\documentclass[%
 reprint,
 amsmath,amssymb,
 aps,
]{revtex4-1}

\usepackage{graphicx}
\usepackage{psfrag}
\usepackage{dcolumn}
\usepackage{bm}

\usepackage{hyperref} 
\usepackage{xcolor}
\hypersetup{
    colorlinks,
    linkcolor={red!50!black},
    citecolor={blue!70!black},
    urlcolor={blue!90!black}
}
\begin{document}

\preprint{APS/123-QED}

\title{Nanoscale dynamics during self-organized ion beam patterning of Si:\\I. Ar$^+$ Bombardment}

\author{Peco Myint}
 \email{peco@bu.edu}
\affiliation{%
Division of Materials Science and Engineering, \\Boston University, Boston, MA 02215 USA
\\ \& X-ray Science Division, Argonne National Laboratory, 
\\Lemont, Illinois 60439, USA}

\author{Karl F. Ludwig, Jr.}%
 \email{ludwig@bu.edu}
\affiliation{%
Department of Physics and\\Division of Materials Science and Engineering, \\Boston University, Boston, MA 02215 USA}
\author{Lutz Wiegart}
\author{Yugang Zhang}
\author{Andrei Fluerasu}
\affiliation{
 National Synchrotron Light Source II, \\Brookhaven National Lab, Upton, NY 11973 USA
}%

\author{Xiaozhi Zhang}
\author{Randall L. Headrick}
\affiliation{
 Department of Physics and Materials Science Program, \\University of Vermont, Burlington, VT 05405 USA
}%

\date{\today}

\begin{abstract}
{\noindent X-ray Photon Correlation Spectroscopy (XPCS) is used to investigate the fluctuation dynamics during self-organized nanopatterning of silicon by Ar$^+$ bombardment at 65$^{\circ}$ polar angle.  
Rich structure is observed in the development of the correlation dynamics as seen in the evolving correlation time $\tau(q_{||})$ and fluctuation relaxation exponent $n(q_{||})$.  On length scales of the ripple structure, local structure becomes ever more long-lived as coarsening progresses.  In addition, $\tau(q_{||})$ develops a peak on length scales corresponding to the ripple wavelength.  As patterning progresses, correlation times become asymmetric between the positive and negative directions, suggesting the possibility of different dynamics on the slopes facing toward and away from the ion beam.  Relaxation exponents show evolution from linear dynamics at early times to compressed exponential relaxation at low wavenumbers and stretched exponential relaxation at high wavenumbers.  Compressed exponential behavior is reminiscent of stress relaxation processes observed in glasses.}

\end{abstract}

\maketitle


\section{\label{sec:level1}Introduction}
Broad beam low-energy ion bombardment of surfaces can lead to the self-organized formation of patterns including nanodots \cite{ozaydin2008effects}, nanoscale ripples \cite{chan2007making} and nanoscale pits/holes \cite{wei2009self}, as well as to ultrasmoothening \cite{moseler2005ultrasmoothness}. The differences in morphology can be achieved by varying irradiation conditions such as ion energy, fluence, bombardment angle, ion species and substrate \cite{cuerno2020perspective}.
In the case of elemental semiconductors patterned at room temperature the surface is amorphized by the ions and off-axis bombardment can produce ripple patterns.  Competing theories of self-organized ion beam nanopatterning advocate for or combine models of different physical processes believed to play important roles, including curvature-dependent sputtering  \cite{sigmund1969theory,sigmund1973mechanism,bradley1988theory}, lateral mass redistribution \cite{carter1996roughening}, surface diffusion \cite{bradley1988theory}, ion-enhanced viscous flow \cite{umbach2001spontaneous} and stress-induced flow \cite{castro2012hydrodynamic,castro2012stress,norris2012stress,moreno2015nonuniversality,munoz2019stress}. Despite much experimental study, agreement on which effects dominate in a given situation has not been reached.  Theory and experiment are often compared using the average \textit{kinetics}, i.e. the evolution of the nanoscale surface structure spatially averaged over the area sampled by the experiment.  Parameters entering theories are often poorly known \textit{a priori} and can therefore be varied to fit the average kinetics observed in a given study, allowing competing theories to claim agreement with experiment.

Going beyond the average \textit{kinetics} to examine the detailed fluctuation \textit{dynamics} of the nanopatterning process offers a new route to test theory and gain better understanding.  By \textit{dynamics} we mean the temporal evolution of fluctuations about the average nanoscale structure.  The dynamics of such surface evolution is becoming accessible through developments in the coherent x-ray scattering technique of X-ray Photon Correlation Spectroscopy (XPCS) \cite{sutton2008review}.  Continued increases in coherent x-ray flux at synchrotron and free-electron laser sources is enabling the application of XPCS to surface growth and patterning, but such studies remain few in number and the technique's full potential is not yet clear.  Thus, an important part of the current work is also the continued exploration and development of the technique's capabilities for such investigations.  XPCS studies of ion beam nanopatterning have previously been performed for Ar$^+$ patterning of GaSb \cite{bikondoa2013ageing} and of SiO$_2$ \cite{mokhtarzadeh2019nanoscale}.  In both cases, however, the dominant length scale in the system - that of the ripples formed on the surface - coarsens beyond the length scales that could be observed in the experiment.  In the present study (Part I) and its companion paper (Part II \cite{myint2020nanoscale}) we examine the classic cases of Si surface nanopatterning by Ar$^+$ and Kr$^+$ respectively.  We are able to examine dynamics on the length scales of the self-organized ripples, and to compare those results with predictions of theory.  

In the nonlinear regime, theories of surface evolution cannot be solved analytically and simulation must be used to determine their predictions. Moreover, obtaining sufficient statistics for analysis of the dynamics requires more lattice realizations than for analyzing average kinetics.  Therefore, within the scope of the present paper we could compare results with predictions of a single theory - a recent minimal model by Harrison, Pearson and Bradley (HPB) which incorporates a cubic nonlinearity \cite{pearson2014theory,harrison2017emergence}. It's known that the full inclusion of lower order terms to the widely-used anisotropic Kuramoto-Sivashinsky (aKS) equation \cite{makeev2002morphology} can also give several kinetics behaviors similar to those observed below, including interrupted coarsening, broken parity, and development of ripples with preferential asymmetric slopes \cite{loew2019effect}.  Therefore, our focus on simulations of the minimal HPB model does not imply the inability of other models to produce similar results.  

The plan of the paper is as follows: In Sect. II below we describe the methods used in the experiments and simulations.  Section III provides a broad brush overview of the basic behavior observed in the speckle-averaged x-ray intensity evolution during nano-ripple formation.  The early stages of the patterning process are analyzed in Sect. IV to determine the linear theory coefficients which can themselves be compared with theoretical predictions and which inform the parameters used in the subsequent simulations.  Section V examines the late-stage coarsening kinetics in both experiment and simulation while the evolution of fluctuation dynamics is examined in Sect. VI.  The results and their implications are discussed in Sect. VII.

\section{Methods}

\begin{figure}
\includegraphics[width=3.2 in]{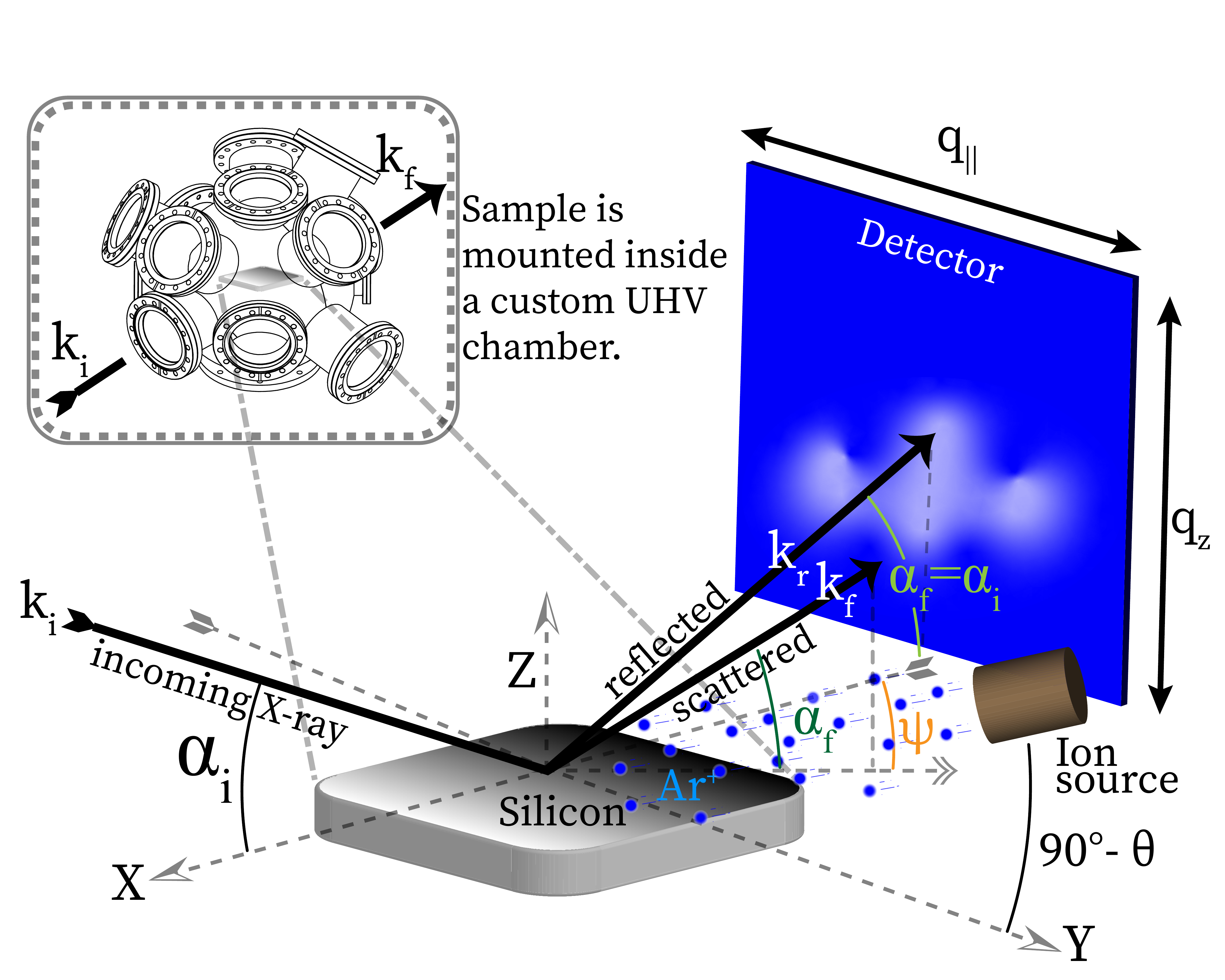}%
\caption{\label{fig:GISAXS} Schematic diagram of the GISAXS experiment. The ion source is placed at the polar angle $\theta$, which causes self-organized rippling on the silicon surface. The sample is positioned so that the X-ray incident angle $\alpha_i$ is slightly above the critical angle of total external reflection. The scattering is recorded as a function of the exit angles $\alpha_f$  and $\psi$ using a 2D detector.}
\end{figure}
\begin{figure}
\includegraphics[width=3.2 in]{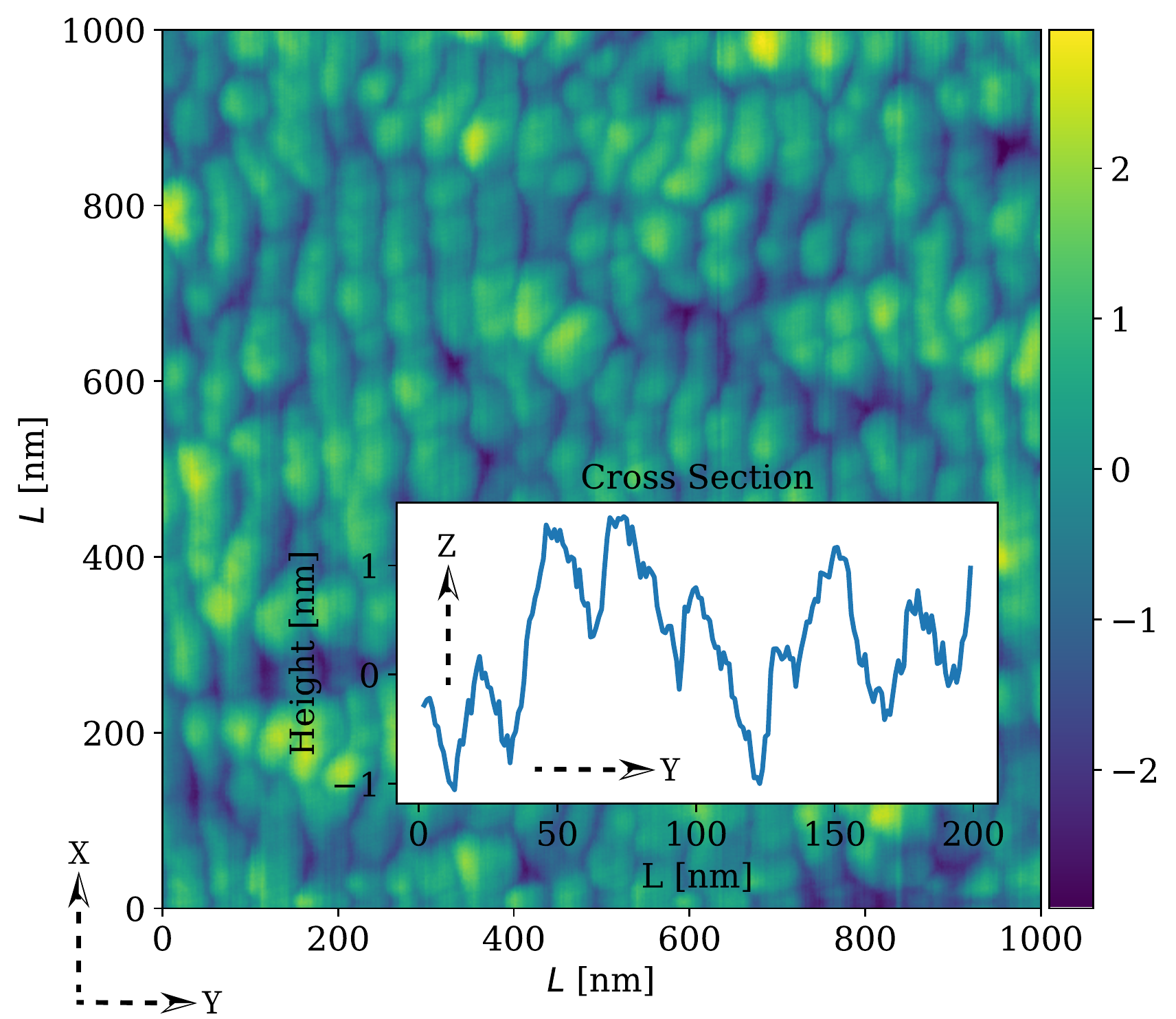}%

\hspace{0.3in}
\includegraphics[width=2.8 in]{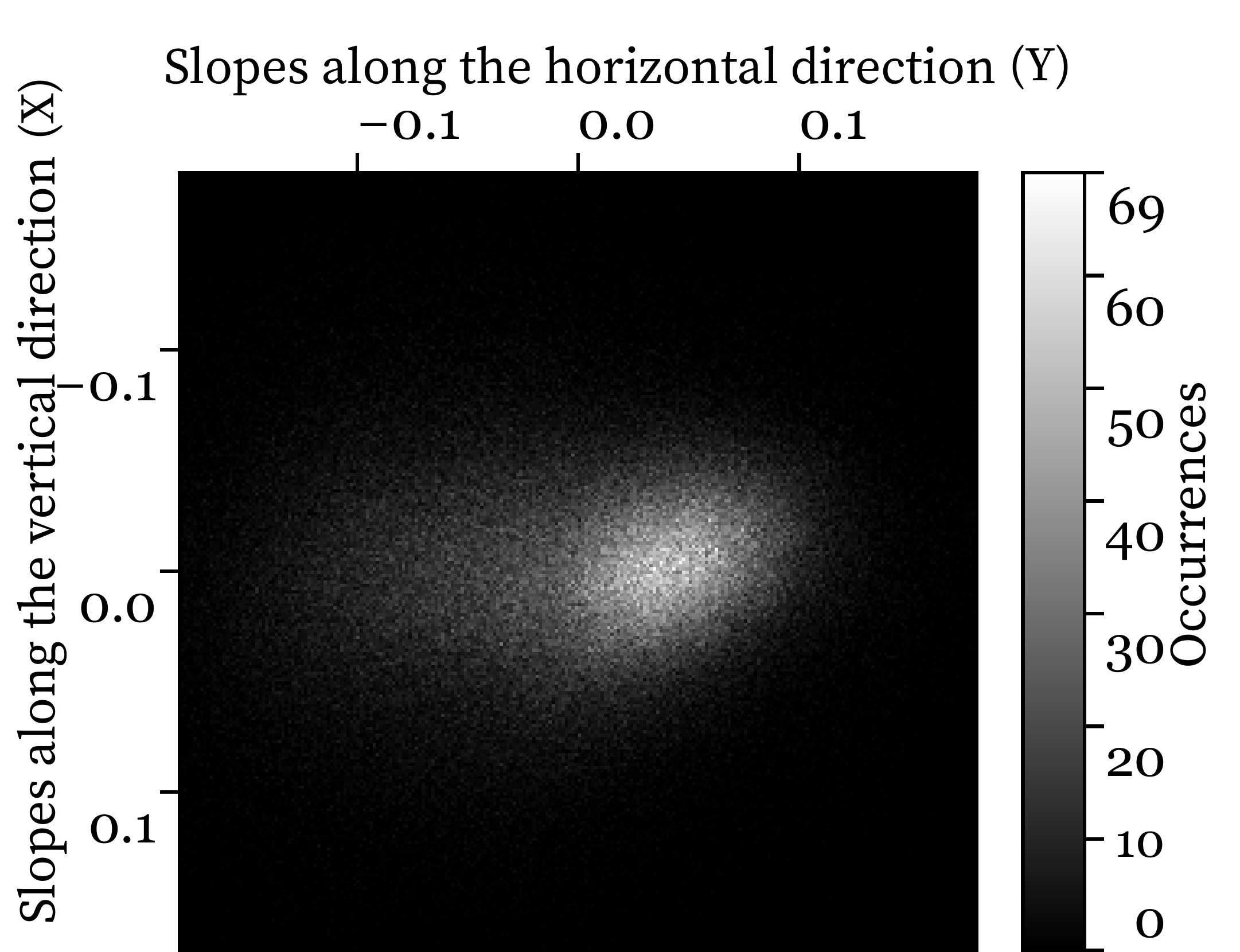}%
\hspace{0.4in}
\caption{\label{fig:AFM}Top: \textit{Post facto} AFM images of silicon surface. The direction of the projection of ion beam's path onto the images is from right to left, while that of X-ray's path is from bottom to top. Bottom: The slope distribution calculated from the AFM image above; a denser distribution of positive slopes indicates that the slopes on the left side of the terraces, shown in the cross-section images, are more defined than the slopes of the other side.}
\end{figure}

\begin{figure*}
\includegraphics[width=3.2 in]{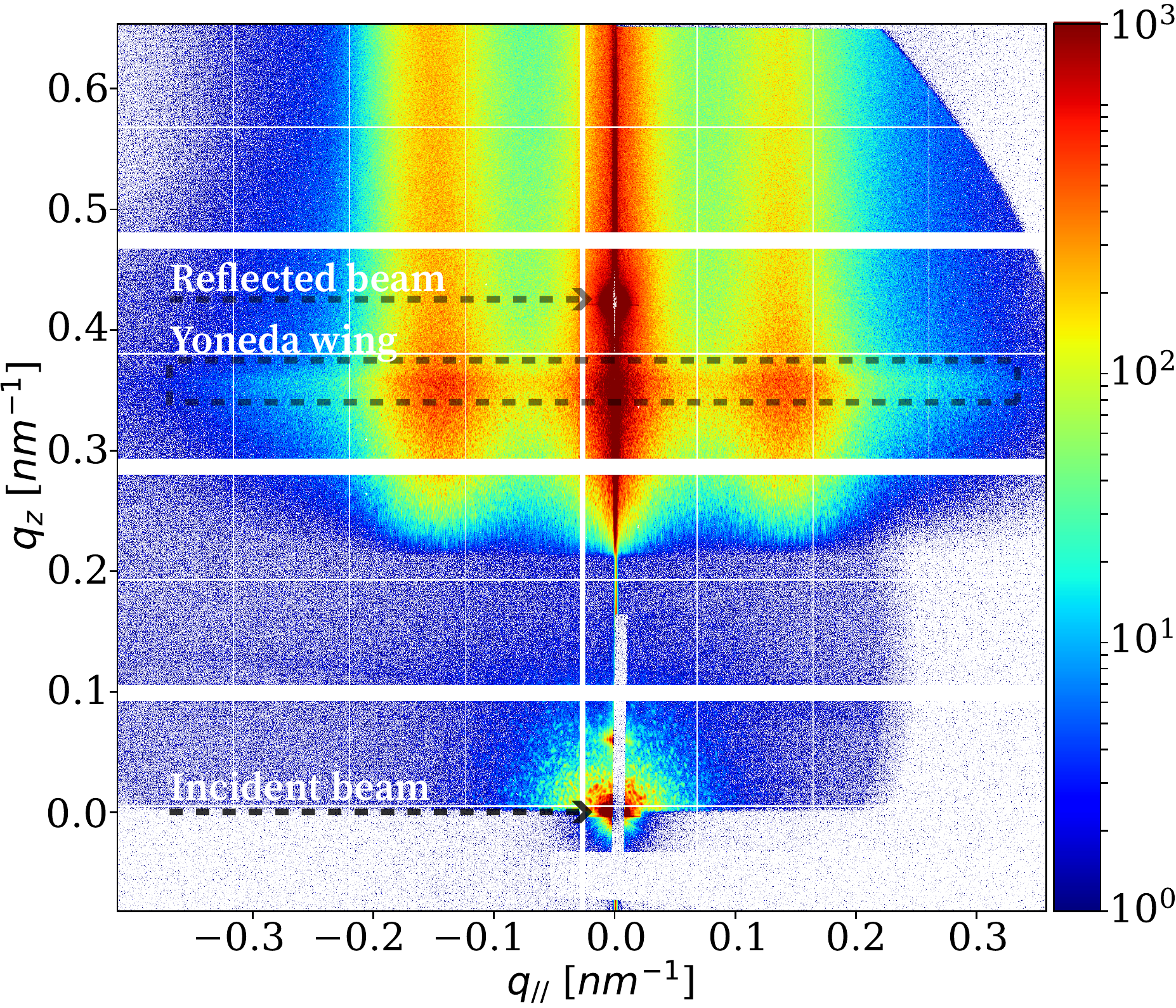}%
\includegraphics[width=3.2 in]{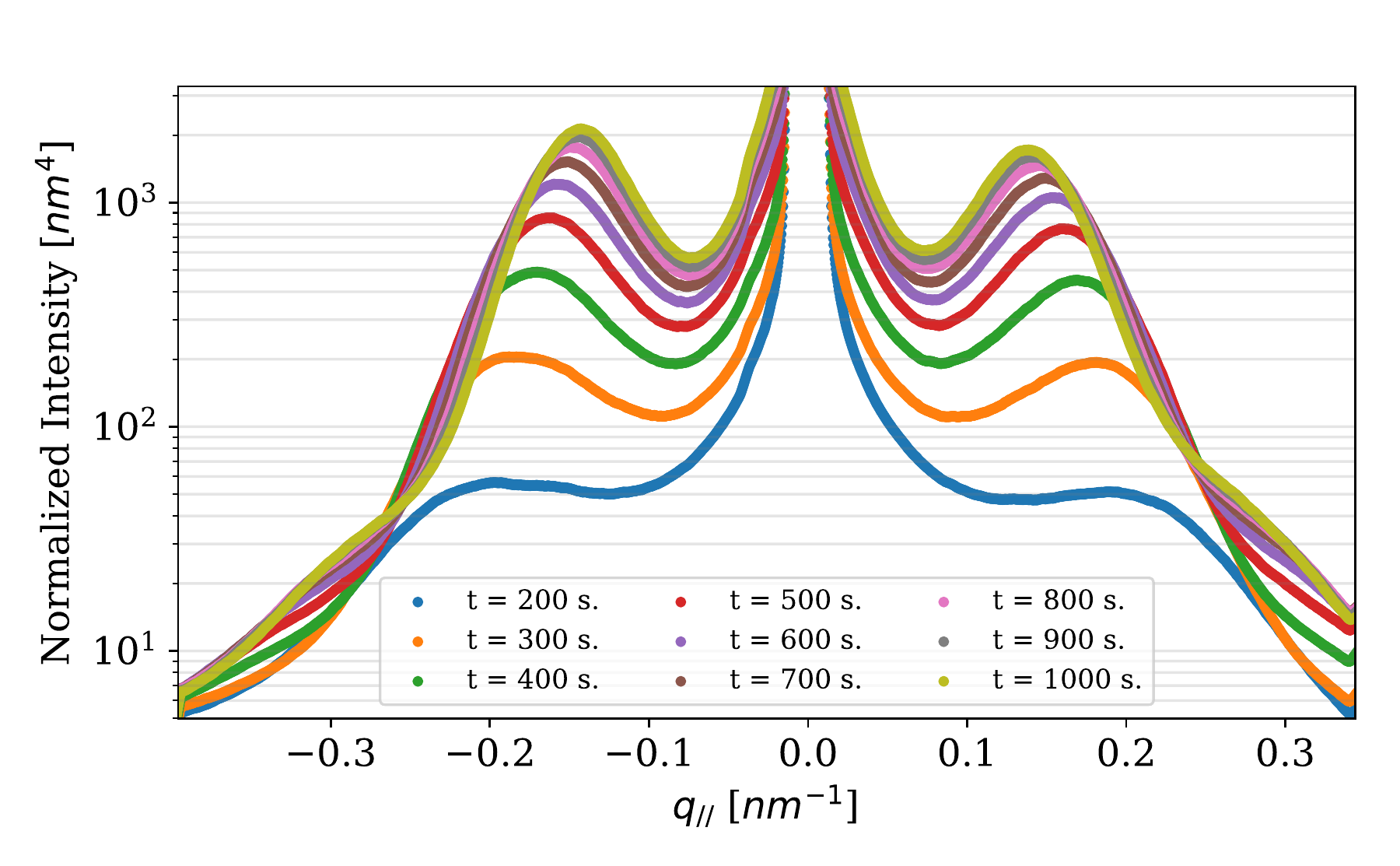}%
\caption{\label{fig:sq_Ar} Left: A detector image during nanopatterning. The Yoneda wing, spread across $q_{z}\,=\,0.36 \; \mathrm{nm}^{-1}$ (corresponding to $q_{z}^\prime\,=\,0.156 \; \mathrm{nm}^{-1}$) is the surface-sensitive scattering exiting the sample at the critical angle, $\alpha_c$. Correlation peaks at $q_{||}\,\simeq\pm\,0.18 \; \mathrm{nm}^{-1}$ are due to the correlated nanoripples on the surface. Right: Evolution of the GISAXS pattern . The 1-d patterns are obtained by averaging along $q_{z}$ across the Yoneda wing, as indicated by the dashed box in the left diagram.}
\end{figure*}

\subsection{\label{sec:Exp-setup}Samples and Ion Bombardment}
The experiments utilized 640 $\mu$m thick p-doped (B) Si(100) wafers cut into 1 $\times$ 1  cm$^2$ pieces and cleaned with acetone, isopropyl alcohol, and methanol.  Samples were firmly affixed to a stage by Indium bonding. To prevent sputtering of impurities onto the surface, the sample stage geometry was designed to ensure that nothing was above the sample surface. The temperature of the water-cooled sample stage was monitored using a thermocouple and the stage was electrically isolated except for a wire leading out to an ammeter in order measure ion flux. The sample holder was mounted in a custom UHV chamber with mica X-ray windows and a base pressure of 5 $\times$ $10^{-7}$ Torr. Samples were kept at room temperature and bombarded with a broad beam of 1 keV Ar$^+$ ions, which were generated by a 3-cm graphite-grid ion source from Veeco Instruments Inc. placed at 65$^{\circ}$ ion incidence angle ($\theta$), as indicated in Fig. \ref{fig:GISAXS}. This ion incidence angle was chosen because it is known to cause self-organized rippling on the silicon surface \cite{madi2011mass,norris2017distinguishing}.  The ion beam flux was measured to be 1 $\times$ 10$^{15}$ ions cm$^{-2}$s$^{-1}$ at the operating chamber pressure of 1 $\times$ $10^{-4}$ Torr. The final fluence was 2.2 $\times$ 10$^{18}$ ions cm$^{-2}$.  The ion beam was sufficiently broad that it uniformly covered the entire sample.

It is important to note that the coordinate system convention of Fig. \ref{fig:GISAXS} follows that often used for GISAXS experiments \textit{and is therefore rotated 90$^{\circ}$ with respect to the coordinate system typically used in the ion bombardment literature}.  Thus, in these experiments "parallel-mode" ripples form with their wavevector pointing in the y-direction rather than in the x-direction, as would conventionally be the situation in studies of ion beam nanopatterning.

\subsection{\label{sec:coGISAXS} Coherent grazing-incidence small-angle X-ray scattering (Co-GISAXS)}
Real-time X-ray scattering experiments were performed at the Coherent Hard X-ray (CHX) beamline at the National Synchrotron Light Source-II (NSLS-II) of Brookhaven National Laboratory. The photon energy of 9.65 keV (wavelength $\lambda = 0.1258 \; \mathrm{nm}$) was selected with a flux of approximately 5 $\times$ $10^{11}$ photon s$^{-1}$ and beam dimensions 10 $\times$ 10 $\mu$m$^2$. Experiments used an Eiger-X 4M detector (Dectris) with 75 $\mu$m pixel size, which was located 10.3 m from the sample. The incident X-ray angle $\alpha_i$ was 0.26$^{\circ}$, which is slightly above the critical angle of total external reflection for silicon of 0.186$^{\circ}$. The projected incident X-ray beam direction on the sample was perpendicular to the projected ion beam direction.  This allowed scattering in the GISAXS geometry to probe the dominant direction of ripple formation for the chosen ion bombardment angle.  The diffuse scattering was recorded as a function of the exit angle $\alpha_f$  and $\psi$ using the 2D detector. The change in X-ray wavevector $\mathbf{q}$ can be calculated from those angles:
\begin{equation}
\mathbf{q} = \mathbf{k_f}-\mathbf{k_i} = 
\begin{pmatrix}
  q_x  \\
  q_y  \\
  q_z 
 \end{pmatrix} = \frac{2\pi}{\lambda}
 \begin{pmatrix}
  \cos (\alpha_i)- \cos (\alpha_f) \cos (\psi) \\
  \cos (\alpha_f) \sin (\psi)  \\
  \sin (\alpha_i) + \sin (\alpha_f) 
 \end{pmatrix}
 \label{equ:wavenumber_conversion}
\end{equation} Since $q_x$ is small, the horizontal component $q_{||}$ (parallel to the surface) can be approximated as simply $q_y$ and the vertical component as $q_z$ (perpendicular to the surface). In the analysis of this paper, we will primarily be interested in the scattering along the Yoneda wing (Fig. \ref{fig:sq_Ar}), which is particularly sensitive to surface structure \cite{renaud2009probing}. For simplicity, we will use the term ``GISAXS pattern'' for the one-dimensional intensity curve $I(q_{||},t)$ obtained by averaging speckles in the detector vertical direction (approximately $q_z$) across the Yoneda wing as shown in Fig. \ref{fig:sq_Ar}.
\subsection{\label{sec:simulations}Simulations}
Simulations were performed using the HPB model \cite{pearson2014theory,harrison2017emergence}:

\begin{eqnarray}
\frac{\partial h(\textbf{r},t)}{\partial t} = &&A \, h_y + S_x \, h_{xx} + S_y \, h_{yy} + \lambda _x \, h_x^2 +\lambda _y \, h_y^2 + \nonumber\\&&\gamma _y \,h_y^3 -B \nabla^{4}h +\eta(\textbf{r},t),
\label{equ:HPB}
\end{eqnarray}
where $\eta(\textbf{r},t)$ is a Gaussian white noise. For $S_x > 0$ and $S_y < 0$, this produces ripples in the $y$-direction.  Numerical integrations were performed on a 2048 $\times$ 2048 lattice using the one-step Euler scheme for the temporal discretization with an integration step $\Delta t$ = 0.001. The spatial derivatives were calculated by standard central finite difference discretization method on a
square lattice with periodic boundary conditions. To check the accuracy of our calculations, we also used the Lam-Shin discretization \cite{lam1998improved} to compute nonlinear terms and found the results were similar. In the simulations, the surface is taken to be initially flat.  For comparison with experiment, the lattice size and time units in the simulation are set as 1 nm and 1 second respectively. 

The linear coefficients $S_y$ and $B$ in the simulation were determined from a preliminary linear theory analysis of the measured early-stage kinetics as discussed in Sect. \ref{sec:early_kinetics}. The $S_x$ coefficient was assigned the same magnitude as $S_y$, but with opposite sign based on the measurements suggested by Norris \textit{et. al} \cite{norris2017distinguishing}. The amplitude of the noise term $<\eta^2>$ was also suggested by the linear theory analysis of kinetics in the early time. The $A$ and nonlinear coefficients were calculated from the sputter yield $Y(\theta)$ curve as discussed in Pearson \textit{et al.},  \cite{pearson2014theory}. In sum, the parameters used in the simulations were: A = -0.26 nm/s, $S_x$ = 0.45 nm$^2$/s, $S_y$ = -0.45 nm$^2$/s, $B$ = 6.96 nm$^4$/s, $\lambda_x$ = 1.94 nm/s, $\lambda_y$ = 1.94 nm/s, $\gamma_y$ = 11.89 nm/s, $<\eta^2>$ = 0.1 nm$^2$/s$^2$, $\Delta t$ = 0.001 s. 
\begin{figure}
\includegraphics[width=3.41 in]{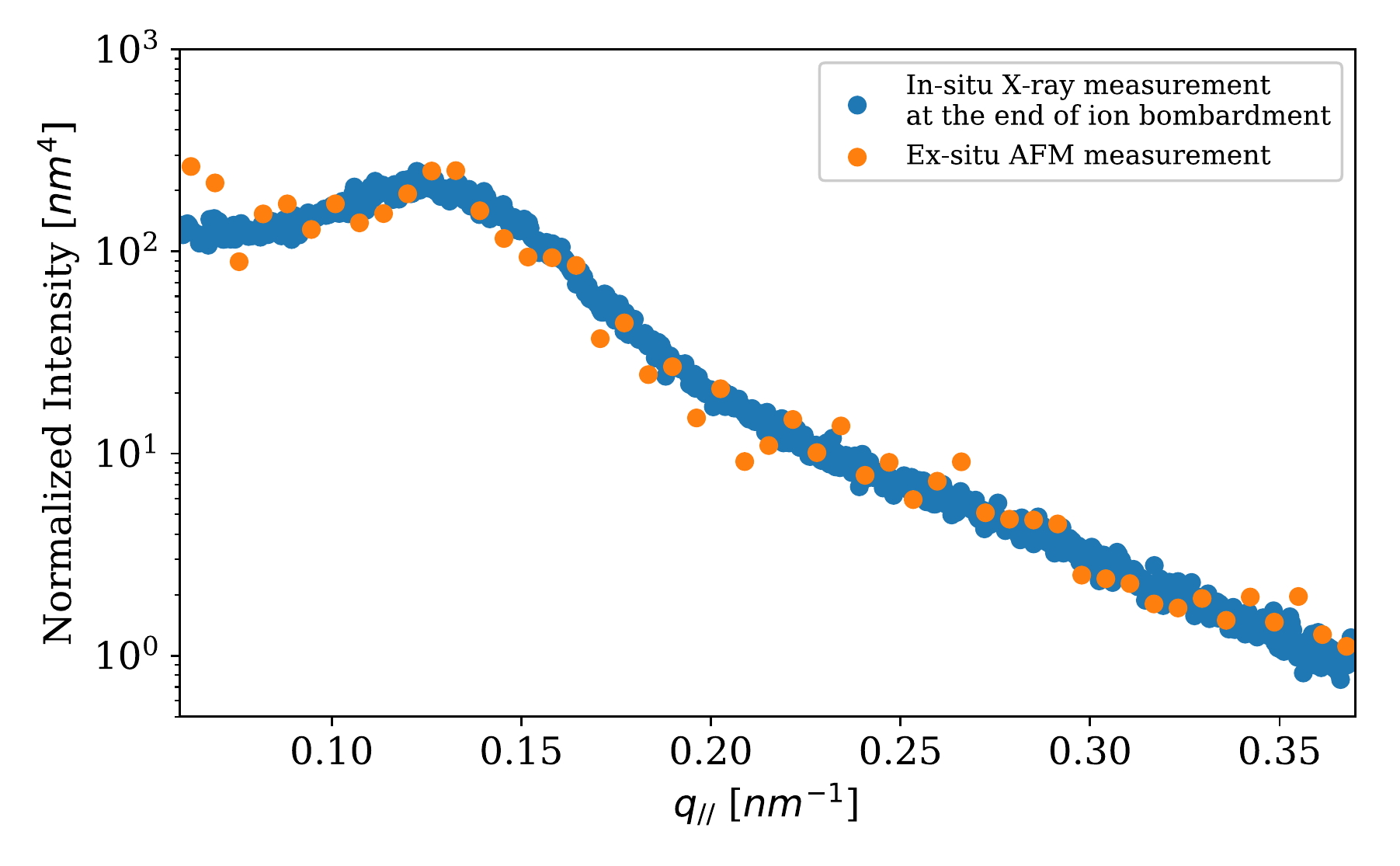}%
\caption{\label{fig:Normalized_intensity} Comparison of final X-ray scattering pattern and predicted GISAXS intensity from the \textit{post facto} AFM topograph.  AFM results are averaged over four points on the sample.}
\end{figure}
For comparison with experiment, lattices were saved after every 1000 steps (i.e. equivalent to every second); the total number of images generated was 1300 and the video is uploaded on Youtube \footnote{HPB equation simulation video:\url{https://youtu.be/JY3n37PR4WI}}.
\subsection{\label{sec:compare_real_Xray}Method of Comparing Real-Space Structure with X-ray Scattering}
In order to connect simulated surfaces and \textit{post facto} Atomic Force Microscopy (AFM) topographs with X-ray scattering, we calculate their predicted GISAXS scattering patterns using the equation:
\begin{equation} 
I(q_x,q_y,q_z) \propto \frac{1}{A} \left | \frac{1}{q_z^\prime} \iint dx \ dy \ e^{-iq_z^\prime h(x,y)}  e^{-i(q_x x+q_y y)} \ \right |^2
\label{equ:Born-approx}
\end{equation}
where $A$ is illuminated area, and $q^\prime_z$, which is calculated by using the refracted incident $\alpha_i' = \sqrt{\alpha_i^2 - \alpha_c^2}$ and exit $\alpha_f' = \sqrt{\alpha_f^2 - \alpha_c^2}$ angles, is the z-component of the wave-vector change inside the material \cite{sinha1988x}. The geometrical value $q_z$ is used for display purposes in detector images since it is zero at the direct beam position on the detector, but in the data analysis, we use $q^\prime_z = 0.156 \; \mathrm{nm}^{-1}$ which is the average $q_z^\prime$ of detector pixels along the Yoneda wing used in the analysis of the X-ray data. In the case of small $q_z^\prime$, the intensity $I(q_{||},t)$ becomes proportional to the height-height structure factor, but for accuracy, the exponential term is kept in the calculations.

The \textit{post facto} AFM topographs show the development of ripple structures (Fig. \ref{fig:AFM}). The GISAXS pattern calculated from the AFM images agrees well with the final GISAXS patterns actually observed as shown in Fig. \ref{fig:Normalized_intensity}.  This allowed the measured GISAXS pattern to be normalized to an absolute scale relative to surface structure height.  Equation \ref{equ:Born-approx} gives the units of intensity as (length)$^4$, and so the resulting units of normalized intensity here are nm$^4$.  This is a natural unit for surface scattering and, when $q_z^\prime$ is small, reflects that the intensity is proportional to the height-height structure factor whose two-dimensional integral in reciprocal space is equal to the square of the RMS roughness.

\section{\label{sec:overview}Overview}

In the experiments, ion bombardment started at $t$ = 0 s, and a clear correlation peak can be seen growing around $t = 100 \; \mathrm{s}$ due to the formation of correlated ripples on the surface.  The initial peak wavenumber is at $q_0 \approx 0.22 \; \mathrm{nm}^{-1}$ so that the initial wavelength of ripples is approximately  $2\pi/$0.22  nm$^{-1}$ $\approx$ 28.6 nm. In Sect. \ref{sec:early_kinetics} below, we quantitatively analyze this behavior using linear theory of nanopatterning. A typical detector pattern and GISAXS intensity patterns at particular times in the evolution are shown in Fig. \ref{fig:sq_Ar}.  After the time regime of linear theory, coarsening occurs with the correlation peak position $\pm q_0$ shifting to smaller wave number. The coarsening proceeds at an ever decreasing rate and, by the end of the experiment, the average GISAXS pattern changes only slowly; the final ripple wavelength suggested by the correlation peak position was approximately $2\pi/$0.12  nm$^{-1}$ $\approx$ 50 nm.  In addition to the primary correlation peak at $\pm q_0$, a harmonic is seen to form at $\pm 2q_0$.  These behaviors are analyzed below.

\section{\label{sec:early_kinetics}SPECKLE-AVERAGED EARLY-TIME KINETICS}
\begin{figure}
\includegraphics[clip,trim={0.0in 0 0 0},width=3.41in]{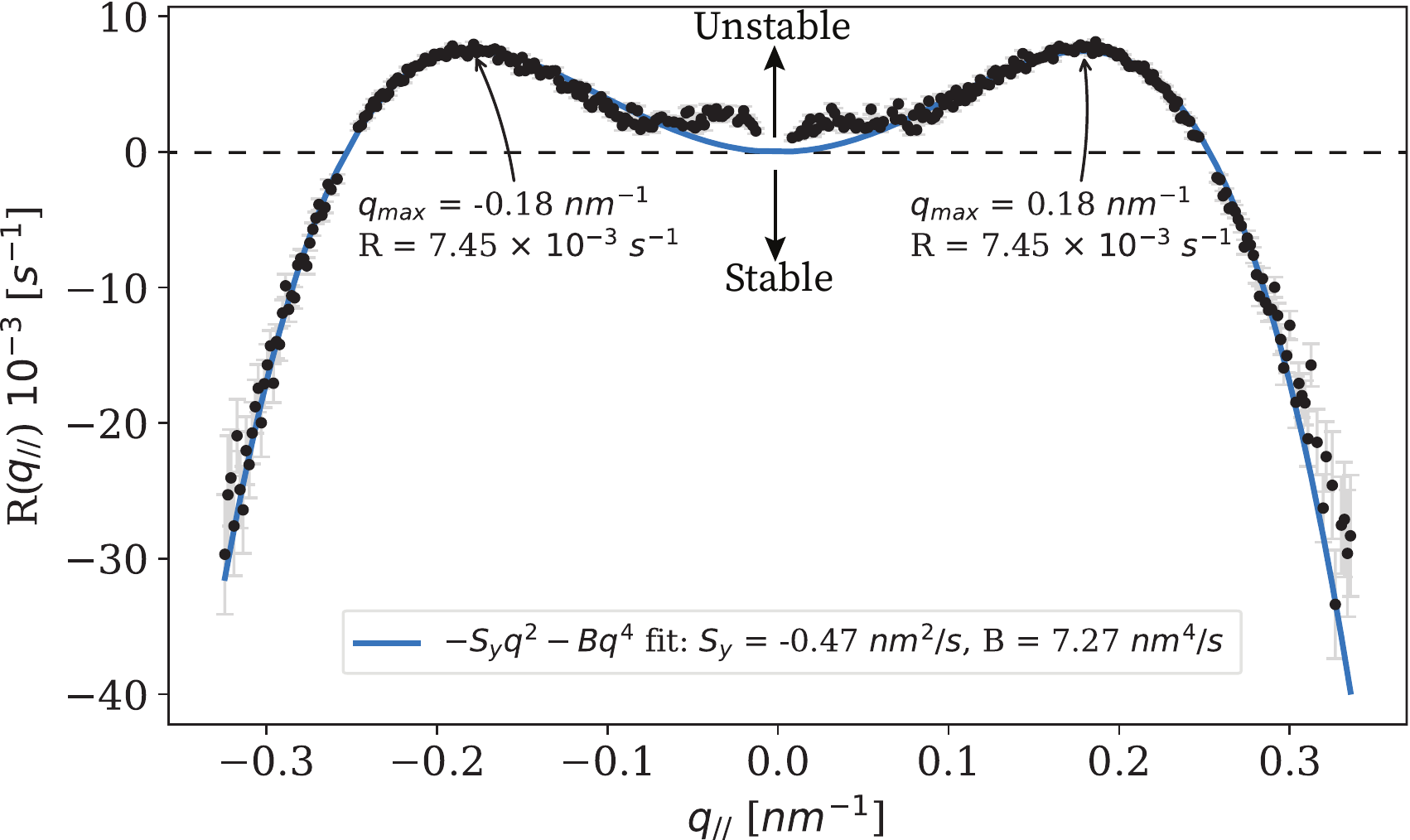}%
\caption{\label{fig:linear-th-fits} Amplification factors obtained from linear theory analysis of speckle-averaged intensity evolution during early stages of nanopatterning.}
\end{figure}
At early stages of nanopatterning, when surface slopes are small, nonlinear equations such as the HPB model reduce to a more tractable linear stability theory which, in reciprocal space, takes the form \cite{bradley1988theory}:
\begin{equation}
\frac{\partial\tilde{h}\left(\mathbf{q},t\right)}{\partial t}=R\left(\mathbf{q}\right)\tilde{h}\left(\mathbf{q},t\right)+\tilde{\eta}\left(\mathbf{q},t\right)
\label{eq: dispersion-general-form}
\end{equation}
where $\tilde{h}\left(\mathbf{q},t\right)$ is the Fourier transform of the surface height $h\left(\mathbf{r},t\right)$, $R\left(\mathbf{q}\right)$ is  the \emph{amplification factor} or \emph{dispersion relation}, and $\tilde{\eta}\left(\mathbf{q},t\right)$ is the Fourier transform of a stochastic noise.  The amplification factor differentiates surface stability or instability; a positive $R(\mathbf{q})$ at a given bombardment angle drives exponential amplification of modes of wavevector $\mathbf{q}$ resulting in surface instability, while a negative $R(\mathbf{q})$ damps fluctuations and stabilizes modes of wavevector $\mathbf{q}$. In the x-ray measurement direction, $R(q)$ is related to the parameters of the nonlinear HPB theory Eq. \ref{equ:HPB} by:
\begin{equation}
R(q_x \approx 0, q_{||}) \equiv R(q_{||}) =-S_{y}\,q_{||}^2-B\,q_{||}^4
\label{equ:long-wave}
\end{equation}
A linear theory analysis of the observed early-time speckle-averaged kinetics thus allows extraction of experimental values for the coefficients $S_y$ and $B$ for comparison with theoretical prediction and for use in the HPB model simulations.  

At early times, when surface roughness is small, the x-ray scattering intensity $I(q,t)$ is proportional to the height-height structure factor, which can be calculated from Eq. 4 to yield \cite{madi2011mass,norris2017distinguishing}:

\begin{eqnarray}
\label{equ:hhstructure-factor}
I(\mathbf{q},t)&&= \left\langle h(\mathbf{q},t) \, h^*(\mathbf{q},t)\right\rangle\nonumber\\&& =\left(I_0(\mathbf{q})+\frac{n}{2R(\mathbf{q})}\right)e^{2R(\mathbf{q})t}-\frac{n}{2R(\mathbf{q})}
\end{eqnarray}
where $n$ is the magnitude of the stochastic noise: $\left\langle \eta\left(\mathbf{r},t\right) \eta\left(\mathbf{r^\prime},t\right) \right\rangle = n \, \delta(\mathbf{r}-\mathbf{r^\prime})\delta(t-t^\prime)$.

To determine $R(q_{||})$, the intensity values $I(q_{||},t)$ were first averaged over 5 detector pixels in the $q_{||}$ direction and 100 pixels in the $q_z$ direction to remove speckle from the scattering pattern.  The temporal evolution of the scattering from each wavenumber bin was then fit with a function of the form $I(q_{||},t) = a(q_{||})  e^{2R(q_{||})t} + b(q_{||})$, with $a$, $b$ and $R$ being the independent fit parameters for each $q_{||}$ bin (Fig. \ref{fig:linear-th-fits}).  The resulting $R(q_{||})$ values are shown in Fig. \ref{fig:linear-th-fits} with subsequent fits to Eq. \ref{equ:long-wave}. The bumps in $R(q_{||})$ at low $q_{||}$'s on each side of the GISAXS pattern are assumed to be due to overlap with tails of the specularly reflected X-ray beam and are not included in the $R(q_{||})$ fitting. Fit values are $S_y = -0.47 \; \mathrm{nm}^2 s^{-1}$ and $B = 7.27 \; \mathrm{nm}^4 s^{-1}$. Nonlinear least square fitting was used for the fits but, since $R(q_{||})$ at a high $q_{||}$ has high error bars, Least Absolute Deviation (LAD) and Ordinary Least Square (OLS) were also examined; they gave similar results. The initial fastest growing wavenumber according to the linear theory is $q^{max}_{||} = \sqrt{|S_y|/(2B)} = 0.18 \; \mathrm{nm}^{-1}$.

The fit values of the curvature coefficient $S_y$ and the ion-induced viscous relaxation coefficient $B$ can be compared with those obtained from fits in previous non-coherent real-time X-ray experiments by our group and collaborators using an ion source with lower fluxes.  Scaled up by the higher ion flux here, Madi \textit{et al.} \cite{madi2011mass} obtained $S_y$ = -1 nm$^2$s$^{-1}$ and $B$ = 5.5 nm$^4$s$^{-1}$.  
Thus the values of $B$ found between the two experiments differ by about 25\% while there is approximately a factor of two difference in the measurements of $S_y$.  This level of agreement/disagreement must be attributed to some combination of different ion sources, with ion flux varied by a factor of 500, and different experimental set-ups.

For theoretical comparison with the measured $S_y$, we examine the erosive formalism of Bradley and Harper \cite{bradley1988theory} and the redistributive formulism of Carter and Vishnyakov \cite{carter1996roughening} in accordance with our use of the HPB model, while acknowledging that stress-driven theories offer competing views \cite{castro2012hydrodynamic,castro2012stress,norris2012stress,moreno2015nonuniversality,munoz2019stress}.  To evaluate the parameters in the erosive and redistributive models we follow the general approaches of Bobes \textit{et al.} \cite{bobes2012ion} and Hofs{\"a}ss \cite{hofsass2014surface} using SDTrimSP \cite{mutzke2019sdtrimsp} binary collision approximation simulations. These give an erosive contribution $S_y^{eros} \approx 0.51$ nm$^2$/s and a redistributive contribution $S_y^{redist} \approx -1.39$ nm$^2$/s, for a total $S_y^{eros+redist} \approx -0.88$ nm$^2$/s.  This splits the difference between the measurement of Madi \textit{et al.} \cite{madi2011mass} and the present one. On the other hand, a different approach \cite{norris2014pycraters} using the PyCraters Python framework \cite{PyCraters2017} for crater function analysis on the SDTrimSP results gives $S_y^{total} \approx -0.58$ nm$^2$/s, closer to our measured value.

\section{Speckle-Averaged Late-time kinetics and \textit{Post Facto} AFM}
\begin{figure}
\includegraphics[width=3.2 in]{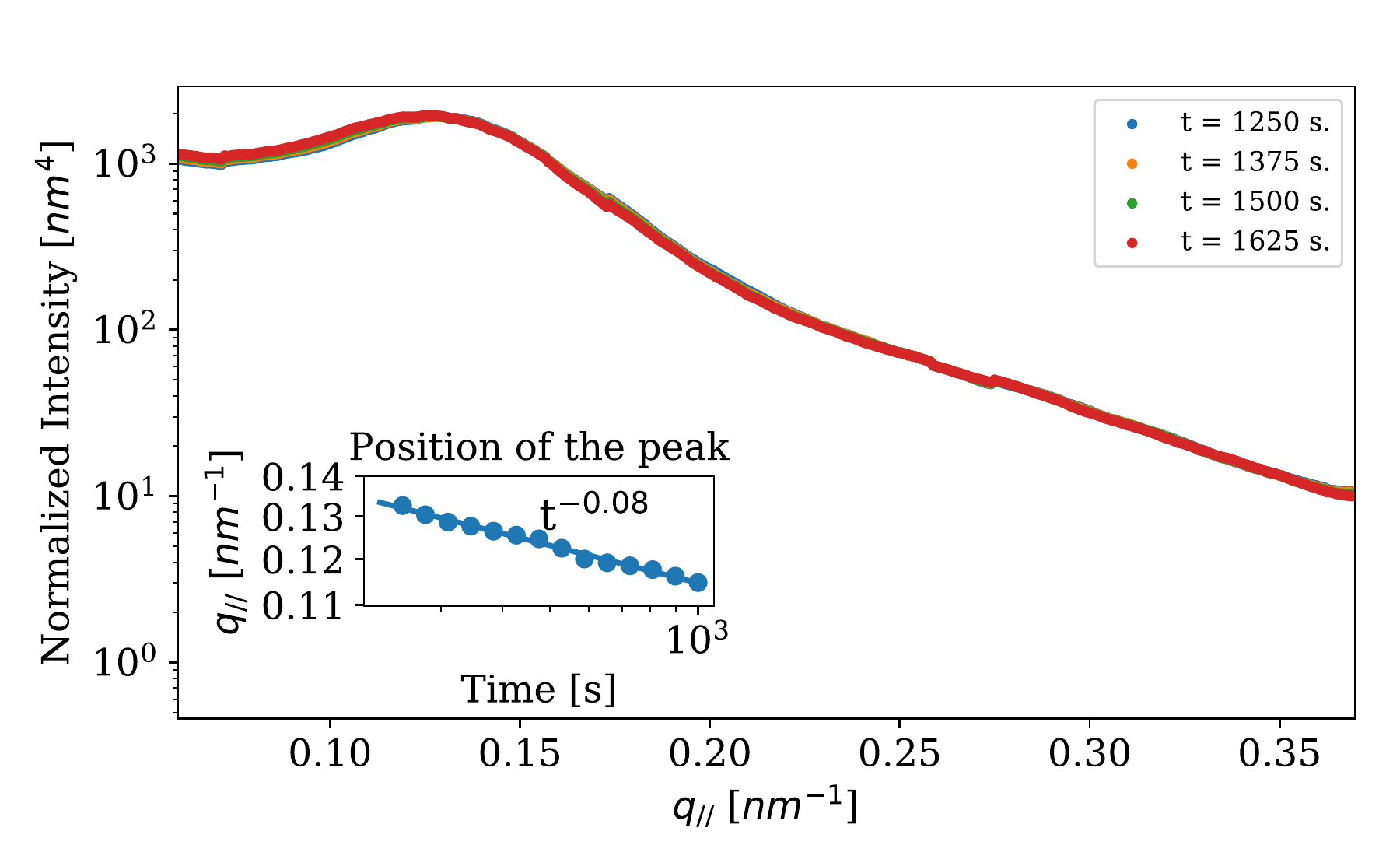}%
\caption{\label{fig:late_kinetics} Coarsening slows down in the late stage of Ar$^+$ patterning of silicon. The inset shows the evolution of the correlation peak position on a log-log scale.}
\end{figure}
The ripple correlation peaks coarsen with time, but at an ever decreasing rate.  Beyond $t$ = 1000 s, the GISAXS pattern changes very little - the peak moves only a few pixels as shown in Fig. \ref{fig:late_kinetics}. While the range of time scales available is too limited to make a definitive statement about the nature of the relaxation, the peak motion can be fit as a weak power law evolution.

At late times, it's well known that the ripples begin to form asymmetric sawtooth structures.  As a result, the scattering pattern becomes asymmetric \cite{ludwig2002si,perkinson2018sawtooth}.  Here it's observed in Fig. \ref{fig:sq_Ar} that the correlation peak at $- q_0$ grows slightly higher than the one at $+q_0$.  More insight comes from the \textit{post facto} AFM topograph, which shows the asymmetric structure, as evidenced by the cut through the topograph and the slope analysis shown in Fig. \ref{fig:AFM}.  Simple calculations of the scattering expected from a sawtooth structure show that, if the negative terrace slope is larger in magnitude than the positive terrace slope on the structure, the negative $q_{||}$ peak should be higher, as observed. In this case, the negative terrace slope is facing the incoming ion beam.  Such calculations also show that, in this case, the harmonic peak at $+2q_0$ should be higher than the one at $-2q_0$, as is also observed.

\begin{figure}
\includegraphics[width=3.2 in]{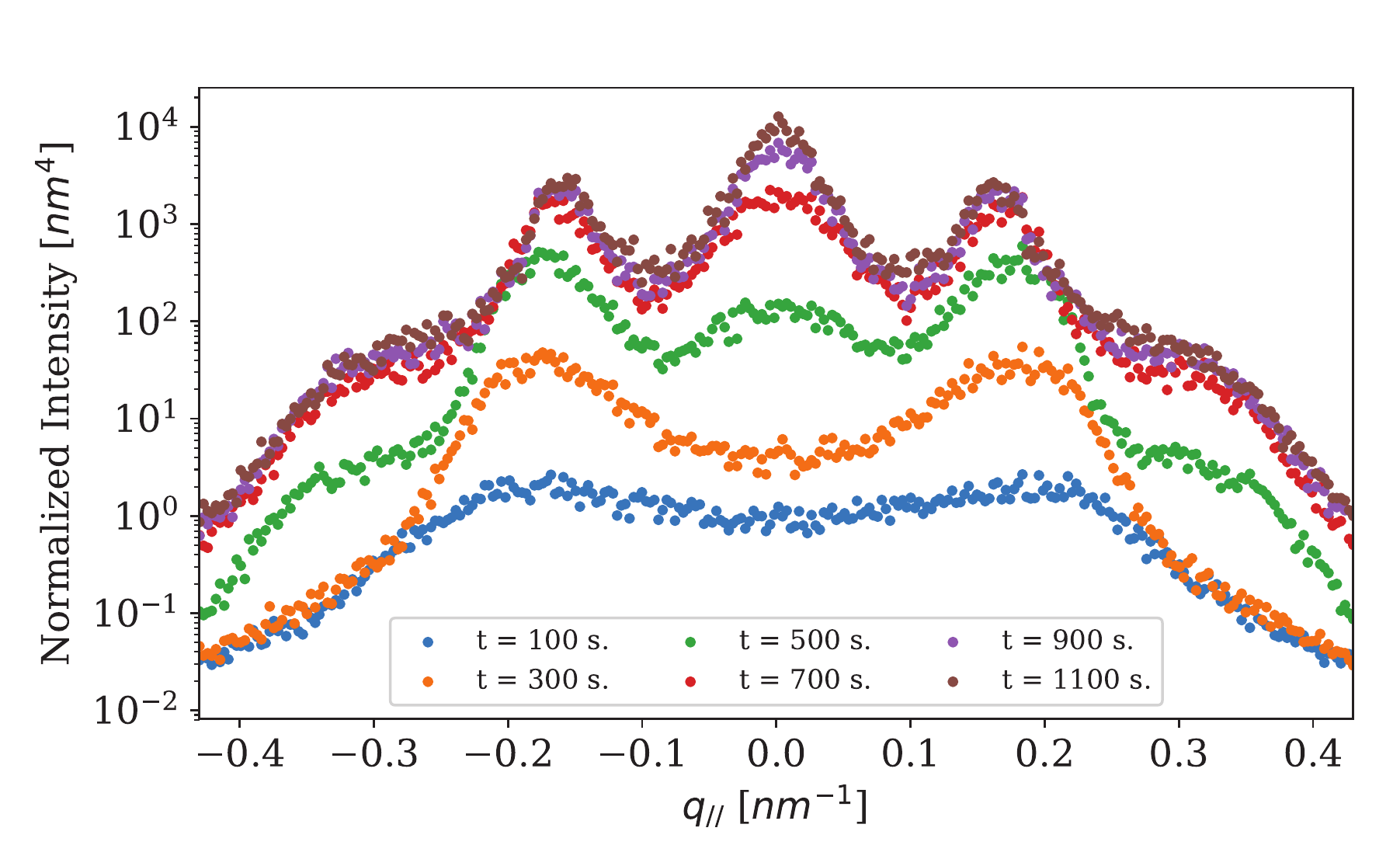}
\caption{\label{fig:Simulation_sq} Simulated GISAXS pattern evolution calculated by averaging results of 100 simulations. As in the experiment, coarsening is observed and kinetic processes slow down over time.}
\end{figure}
\begin{figure}
\includegraphics[width=3.2 in]{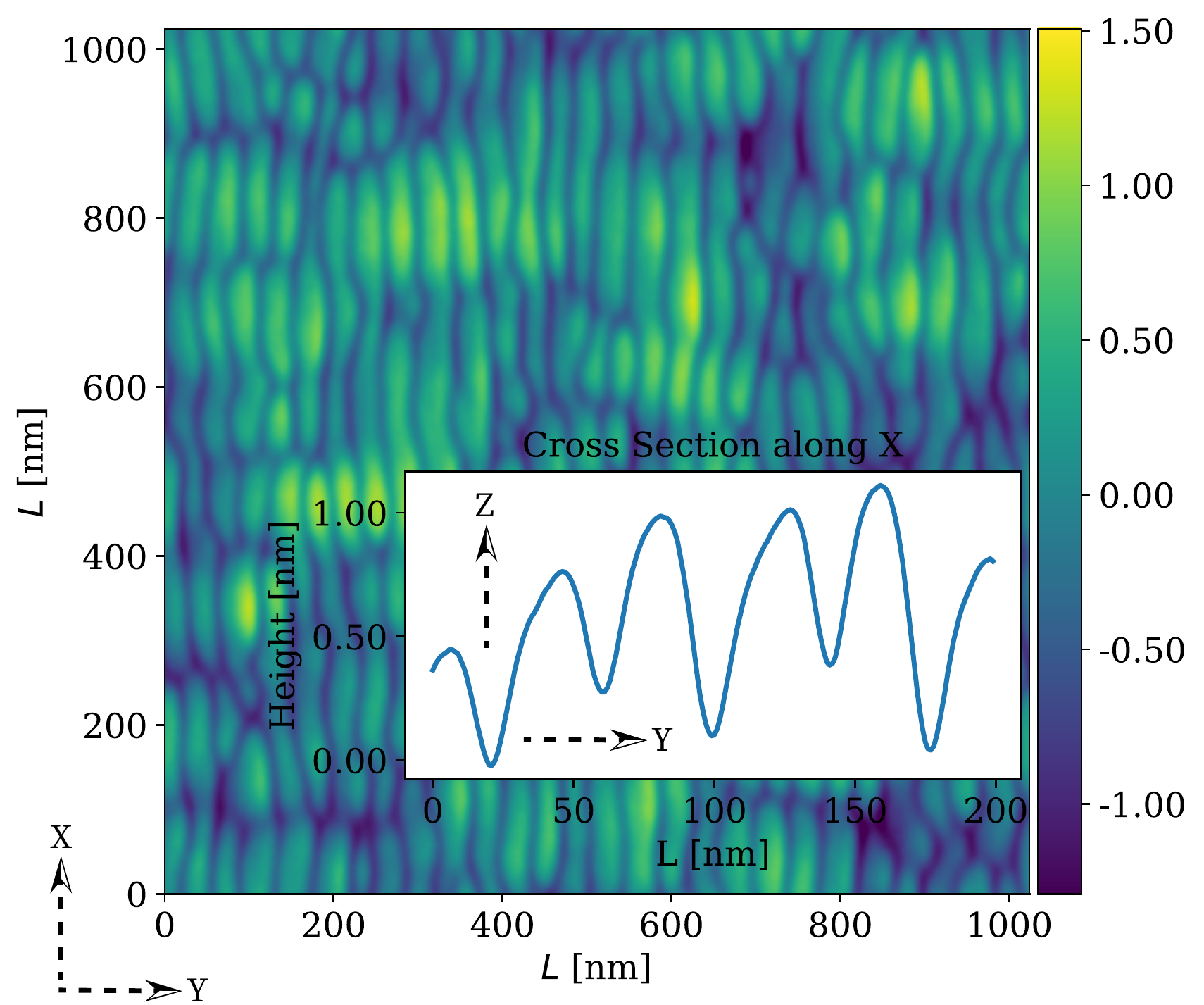}
\includegraphics[width=2.8 in]{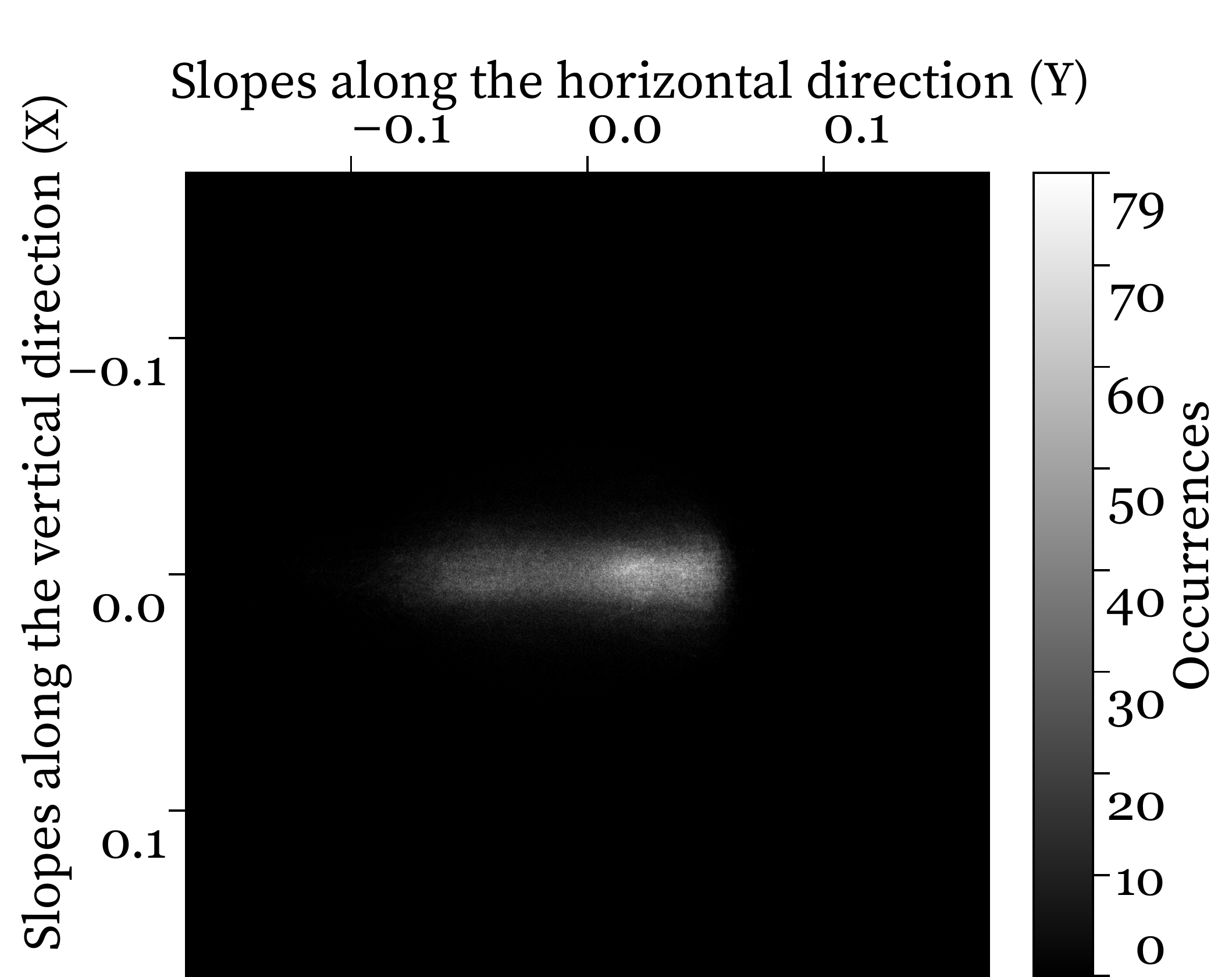}
\caption{\label{fig:Simulation_lattice} Top: A simulated lattice at $t$ = 600 s.
Bottom: Slope distribution calculated from the simulated lattice image. Both results can be compared with the measurements in Fig. \ref{fig:AFM}.}
\end{figure}

The simulations produce speckle-averaged GISAXS scattering patterns (Fig. \ref{fig:Simulation_sq}) showing an initial peak wavenumber $q_0 \approx 0.22$ nm$^{-1}$, in agreement with experiment (Fig. \ref{fig:sq_Ar}), as well as coarsening.  A selected simulation lattice image at $t$ = 1200 s and its slope analysis (Fig. \ref{fig:Simulation_lattice}) can be compared to the \textit {post facto} AFM topograph and slope analysis of Fig. \ref{fig:AFM}.  The maximum time simulated was limited by a subsequent transition to a longer-wavelength sawtooth structure, a phenomenon which has been noted in the literature \cite{gago2002nanopatterning,perkinson2018sawtooth}. The current experiments had not yet reached that regime.
\section{Speckle Correlation Study of fluctuation dynamics}
\begin{figure}
\includegraphics[width=3.2 in]{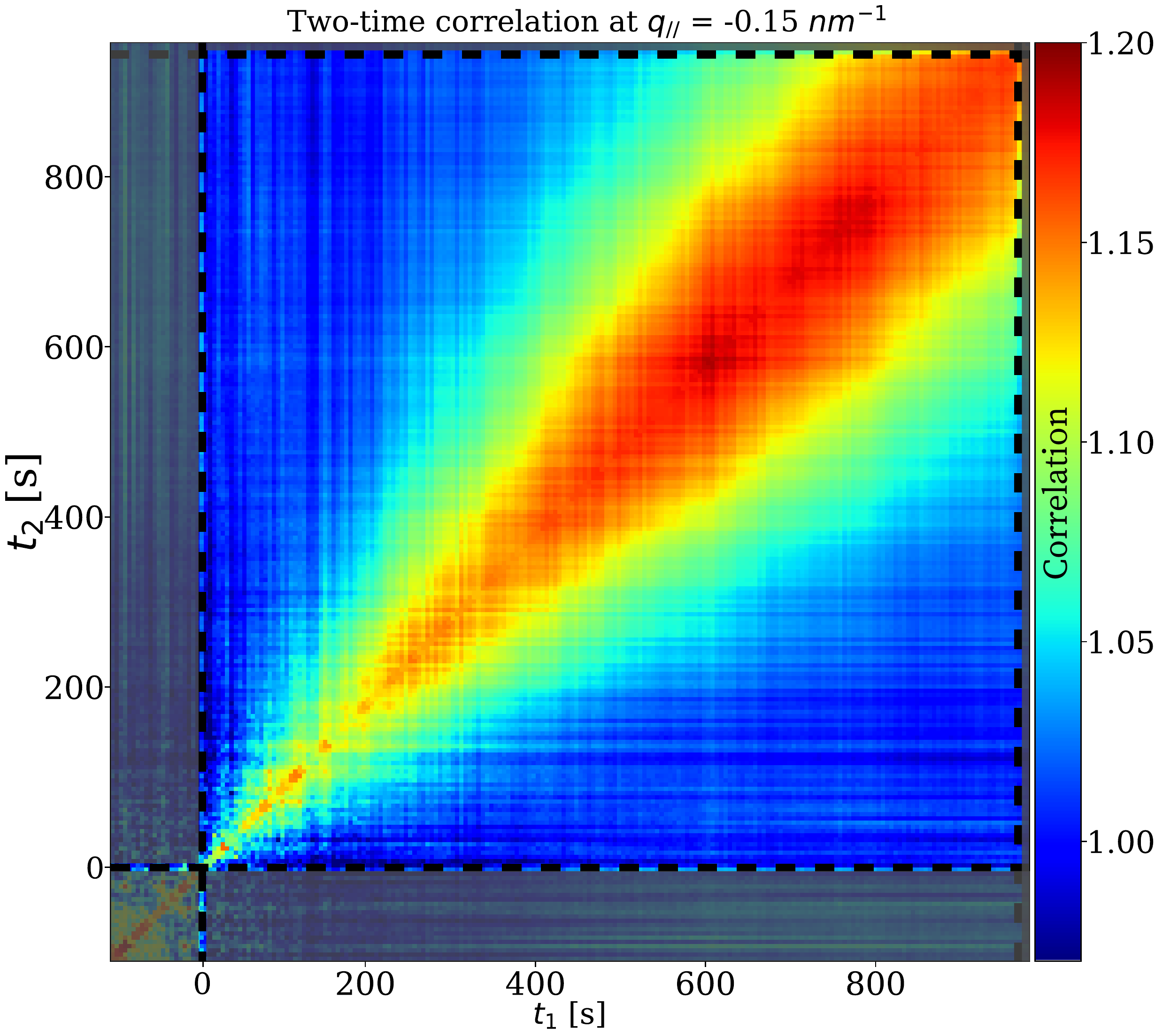}
\caption{\label{fig:TT} Evolution of the two-time correlation function (TTCF). The surface was originally smooth with little scattering, so the function is initially very noisy.  Ion bombardment began at t = 0 s, after 100 s of static scan. The gray areas with dashed boundaries represent the static data taken before and after the ion bombardment.}
\end{figure}
Although the speckle-averaged GISAXS pattern shows the average kinetics, the strength of coherent experiments lies in their ability to measure temporal correlations of the  detailed speckle pattern through XPCS, illuminating the underlying fluctuation dynamics. The two-time correlation function (TTCF) measures how the structure on a given length scale changes between time $t_1$ and time $t_2$ as the sample evolves:
\begin{equation}
C(q_{||},t_1,t_2)= \frac{\left\langle I(q_{||},t_1)I(q_{||},t_2)\right\rangle }{\left\langle I(q_{||},t_1)\right\rangle \left\langle I(q_{||},t_2)\right\rangle} 
\label{equ:twotime}
\end{equation}
where the angular brackets denote an average over equivalent $q_{||}$ values and the denominator values can be considered as speckle-averaged intensities one would have obtained using non-coherent scattering. 

TTCF's are shown in Fig. \ref{fig:TT} for a wavenumber $q_{||}$ near the scattering peak. The central diagonal ridge of correlation going from the bottom left to top right indicates the high correlation expected for $t_1 \approx t_2$. One way to understand how a surface changes on a given length scale is by observing the width of the central correlation ridge, which is a measure of correlation time on the surface. As seen in Fig. \ref{fig:TT}, the width continuously increases, but at a steadily decreasing rate.  At other wavenumbers, the peak width appears to reach a constant value. 
\begin{figure*}
\includegraphics[clip,trim={0.0in 0 0 0},width=3.41in]{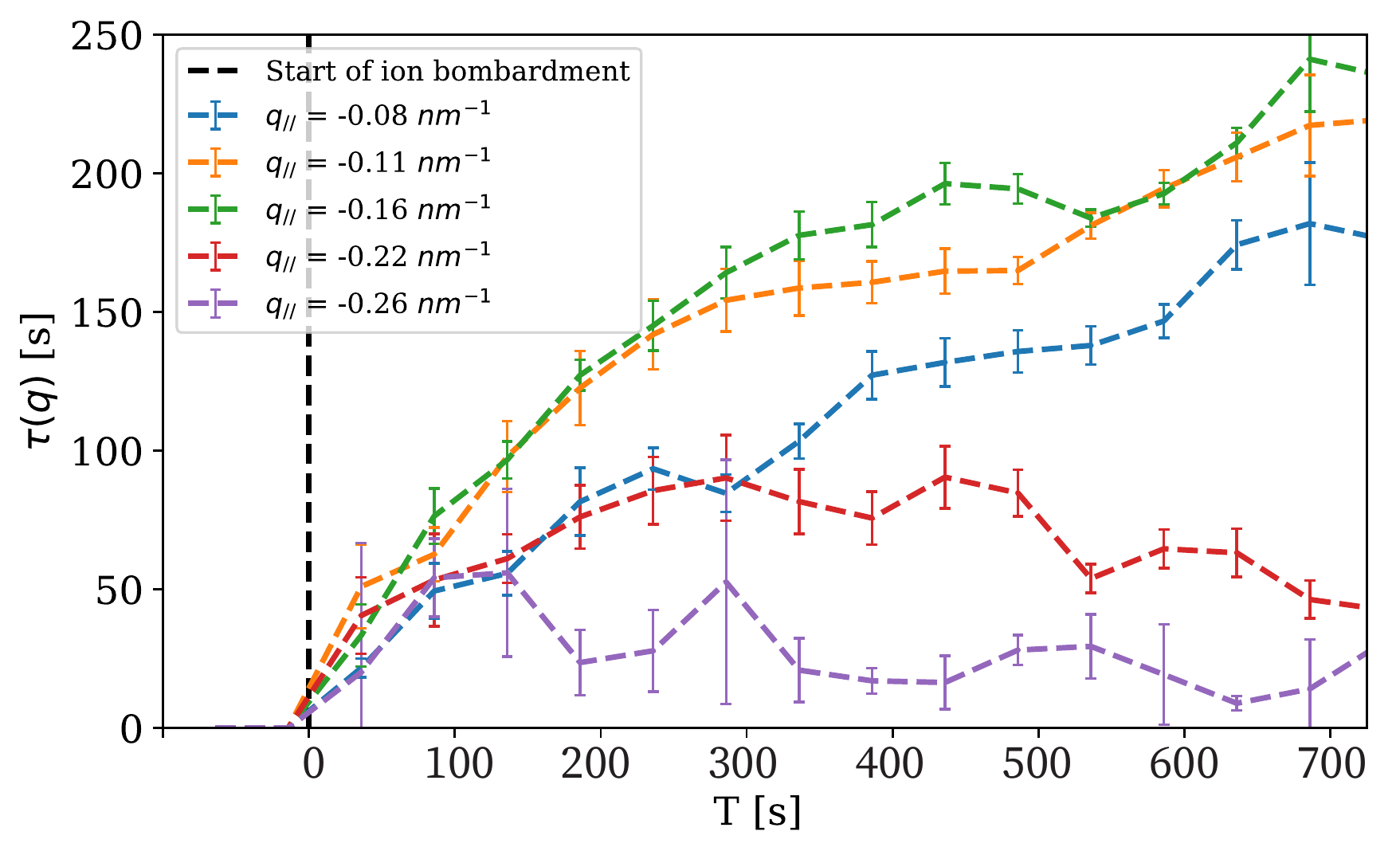}%
\includegraphics[clip,trim={0.0in 0 0 0},width=3.41in]{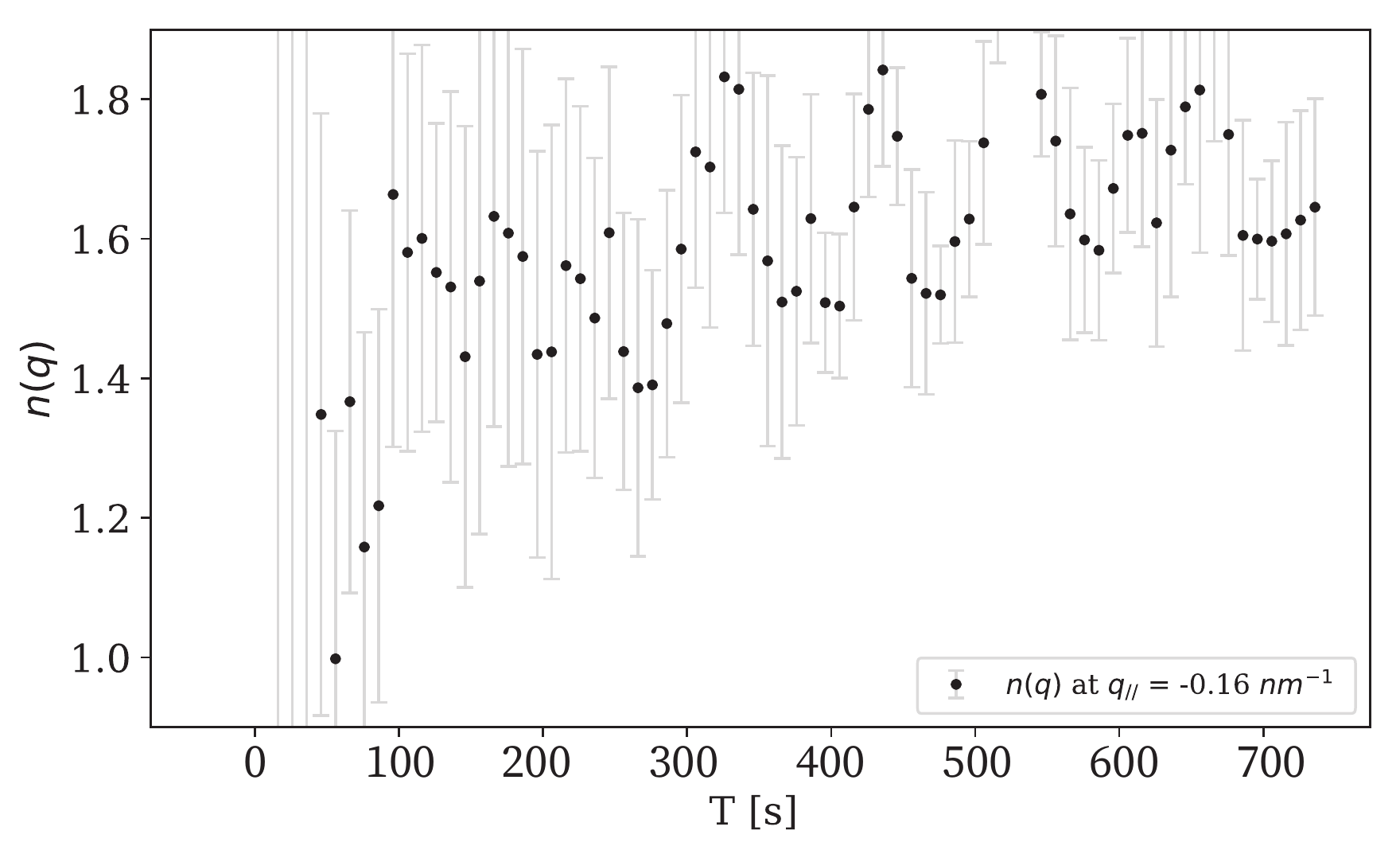}%
\caption{\label{fig:TT_slice_selectq} Evolution of correlation times $\tau(q_{||})$ and relaxation exponents $n(q_{||})$ from KWW fits through diagonal cuts of the TTCF's. Different adjacent time averaging was performed for $n(q_{||})$ to highlight the quick transition from values near 1 to values of 1.6-1.8.}
\end{figure*}
\begin{figure*}[!ht]
\includegraphics[clip,trim={0.0in 0 0 0},width=7.0in]{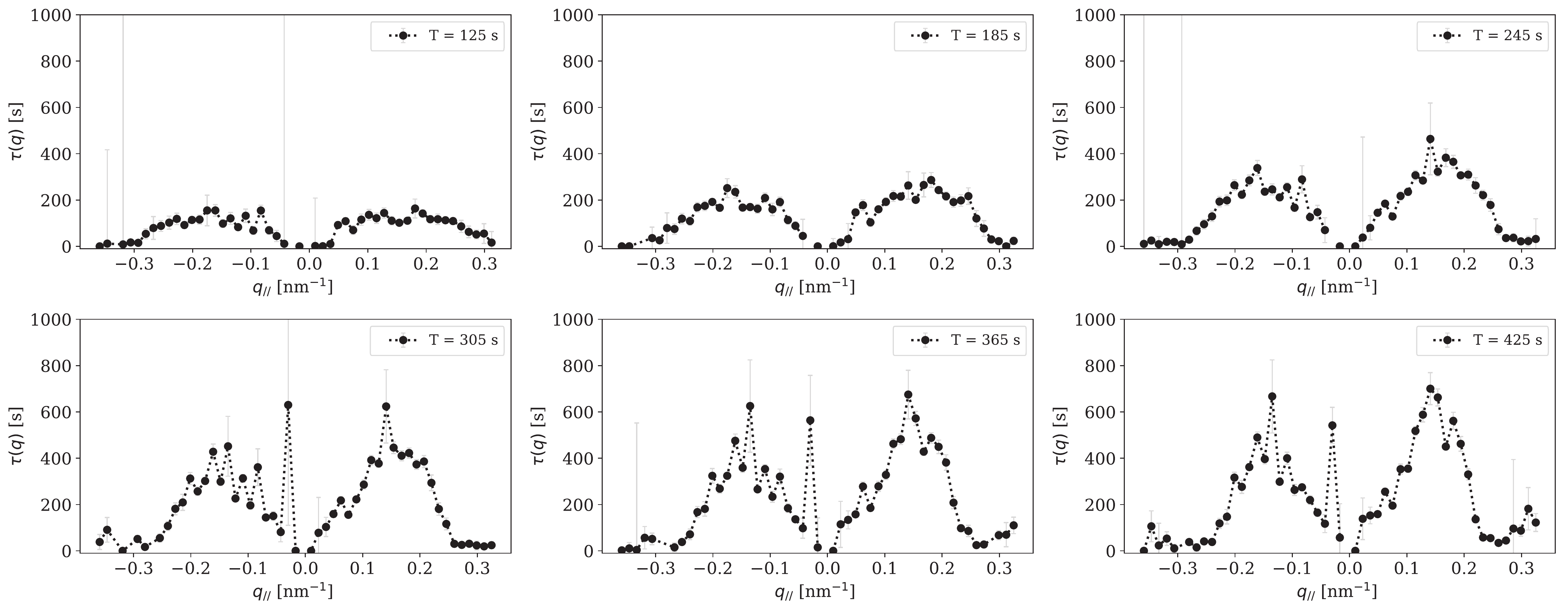}

\includegraphics[clip,trim={0.0in 0 0 0},width=7.0in]{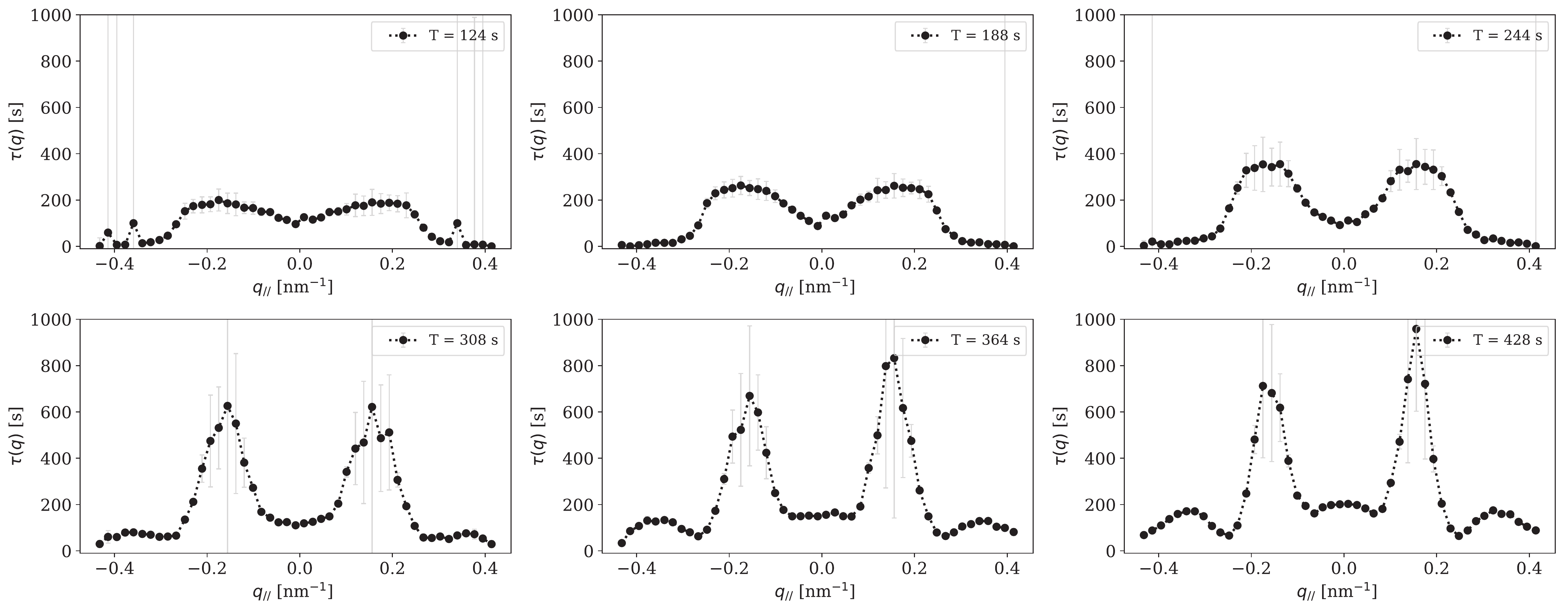}
\caption{\label{fig:TT_slice_allq_Ar} The first two rows are plots of $\tau(q_{||})$ during relatively early stages of patterning obtained from KWW fits of TTCF diagonal cuts. The last two rows are plots of $\tau(q_{||})$ calculated from the HPB simulation.  Note that the experimental results and the simulation results are each plotted to the highest magnitude of $q_{||}$ for which results could reliably be obtained.  Therefore, the horizontal axes are different between the top two and the bottom two rows.}
\end{figure*}
Quantitative measurement of the evolving dynamics is made by taking diagonal cuts through the central ridge at a constant average bombardment time $T = (t_1+t_2)/2 $ as a function of $\Delta t = |t_2 - t_2|$ at each wavenumber $q_{||}$. The decay in correlation with time is fit with the Kohlrausch-Williams-Watts (KWW) form\cite{williams1970non}:
\begin{equation}
g_2^T(q_{||},\Delta t)= b+\beta(q_{||})\, e^{-2({\frac{\Delta t}{\tau(q_{||})}})^{n(q_{||})} },
\label{equ:KWW}
\end{equation}
where $\tau(q_{||})$ is the correlation time, and $n(q_{||})$ is an exponent which determines whether the function is a simple ($n$ = 1), stretched ($0 < n < 1$), or compressed ($n > 1$) exponential. $b$ is the baseline, which was set as 1 or allowed to vary between 0.9 - 1.1. $\beta(q_{||})$ describes the contrast, which depends on experimental factors including the effective resolution of the experiment.  The magnitude of the central diagonal ridge of correlations in Fig. \ref{fig:TT} increases with time, indicating an increasing contrast.  This is probably because background incoherent scattering (e.g. from slits or windows) causes the apparent contrast to decrease at early times when the scattering from the sample is relatively small.  As ripples form on the sample, the scattering from the sample increases and the apparent contrast approaches its limiting value.  Finally, to improve statistics for the fits to Eq. \ref{equ:KWW}, results from $\pm$ 10 s around the central mean growth time $T$ were chosen for averaging.

Figure \ref{fig:TT_slice_selectq} shows the evolution of $\tau$ and $n$ for selected wavenumbers.  Near the peak wavenumber $q_0$, $\tau$ increases continuously, first rapidly and then more slowly.  Away from the peak, the $\tau$ values initially increase but then seem to relax to a steady state.  Near the peak, the relaxation exponent $n$ rapidly increases from approximately one, indicative of simple exponential decay, to 1.6-1.8, showing compressed exponential behavior.

Figure \ref{fig:TT_slice_allq_Ar} shows the behavior of $\tau$ as a function of wavenumber {$q_{||}$} for selected times.  It's seen that the $\tau(q_{||})$ values near the scattering peak $\pm q_0$ grow strongly to become much larger than the relaxation times at smaller and larger wavenumbers.  This distinctive behavior is reproduced in the simulations, as seen in Fig. \ref{fig:TT_slice_allq_Ar}.

Near the end of the experiment, when the correlations are changing more slowly, more detail can be obtained from averaging over a larger time period of $T = 500-1000 $ s, i.e. mean $T = 750$ s, using the auto-correlation function:
\begin{equation}
g_2(q_{||},\Delta t)= \frac{\left\langle I(q_{||},t^{\prime})I(q_{||},t^{\prime}+\Delta t)\right\rangle }{\left\langle I(q_{||}) \right\rangle ^2}.
\label{equ:g2}
\end{equation}
The angular brackets indicate a time averaging over $t^\prime$ and equivalent $q$ values.  Again the calculated $g_2(q_{||},\Delta t)$ function is fit with the KWW form Eq. \ref{equ:KWW}.

\begin{figure*}
\includegraphics[width=3.2 in]{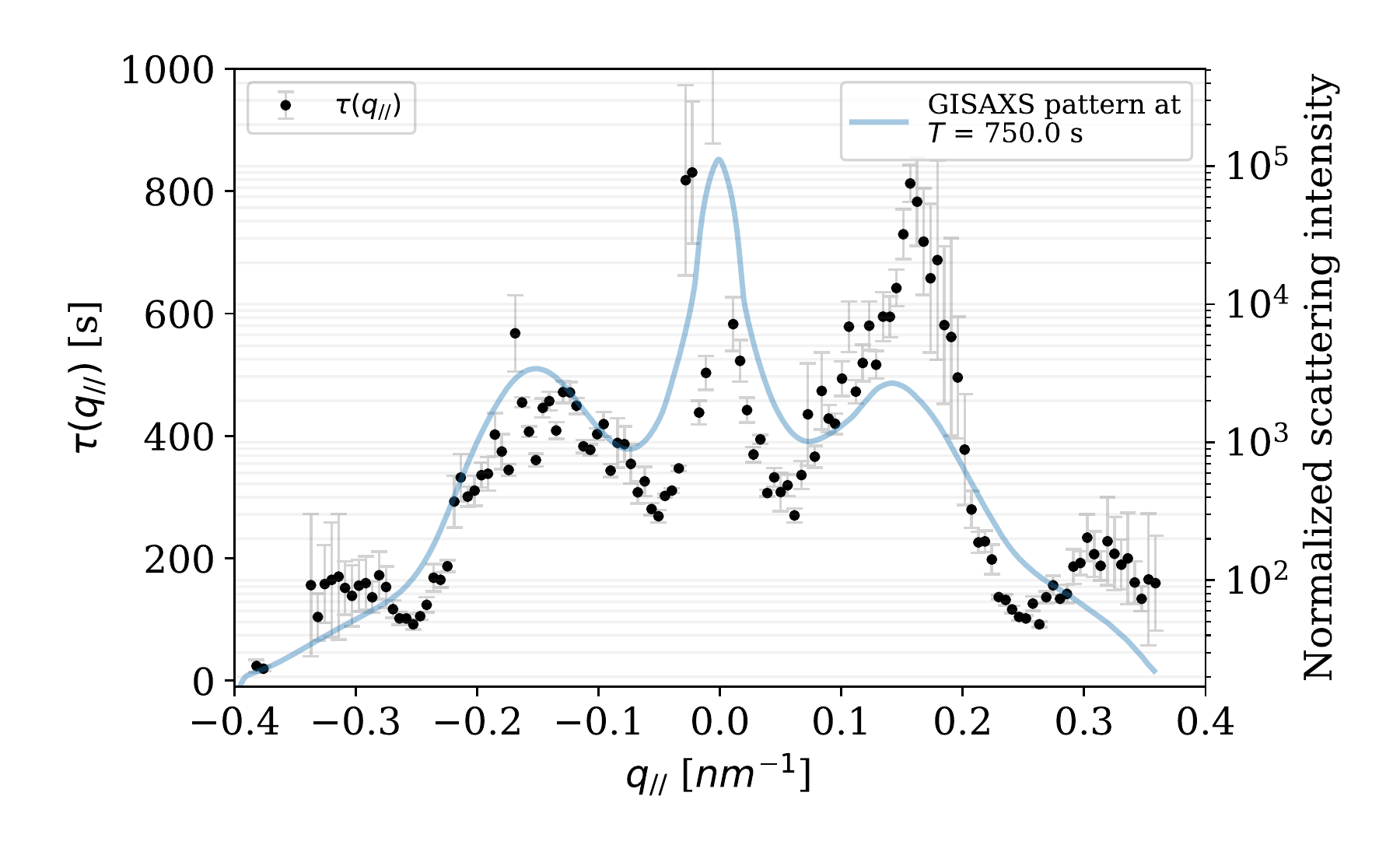}
\includegraphics[width=3.2 in]{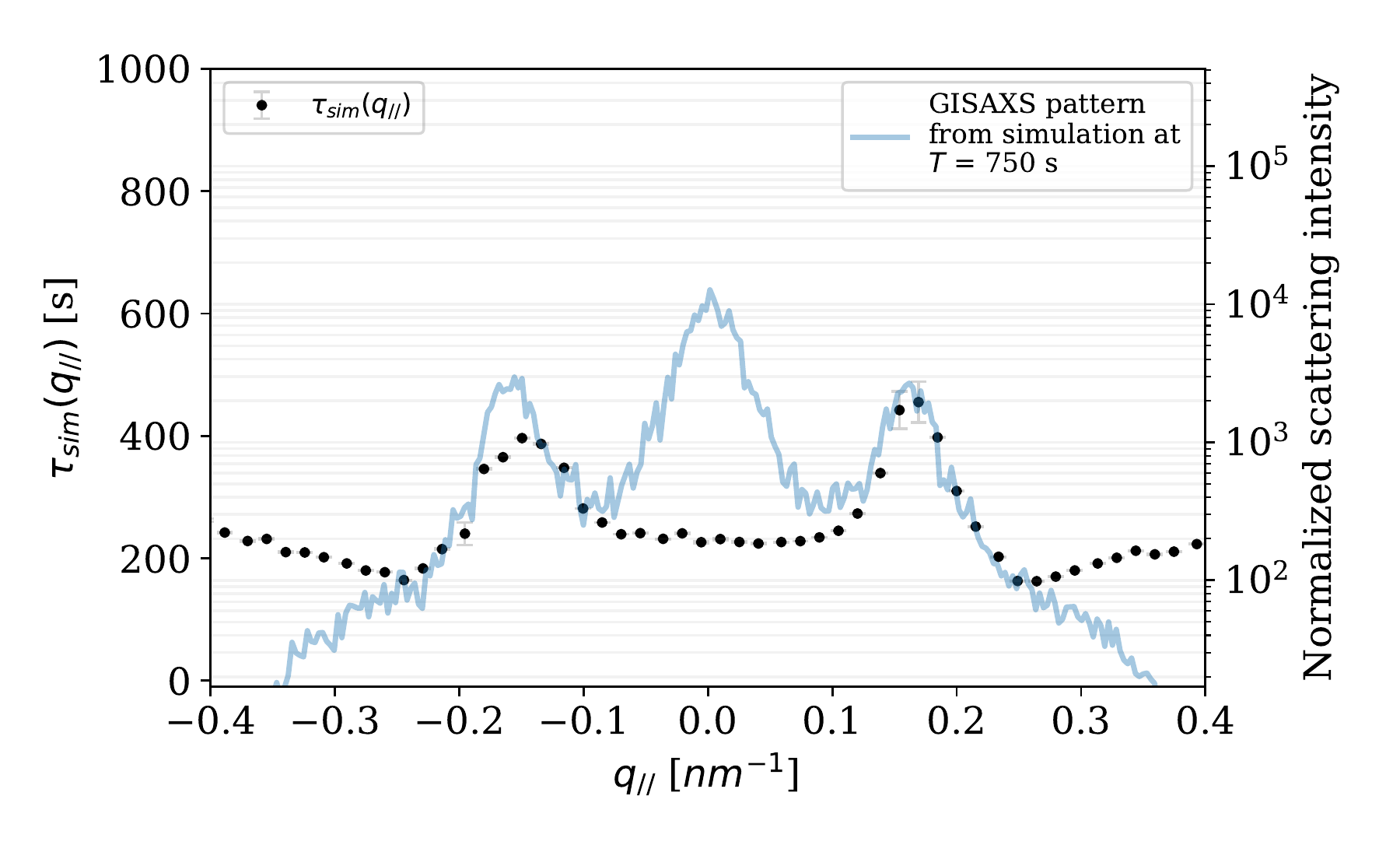}%

\includegraphics[width=3.2 in]{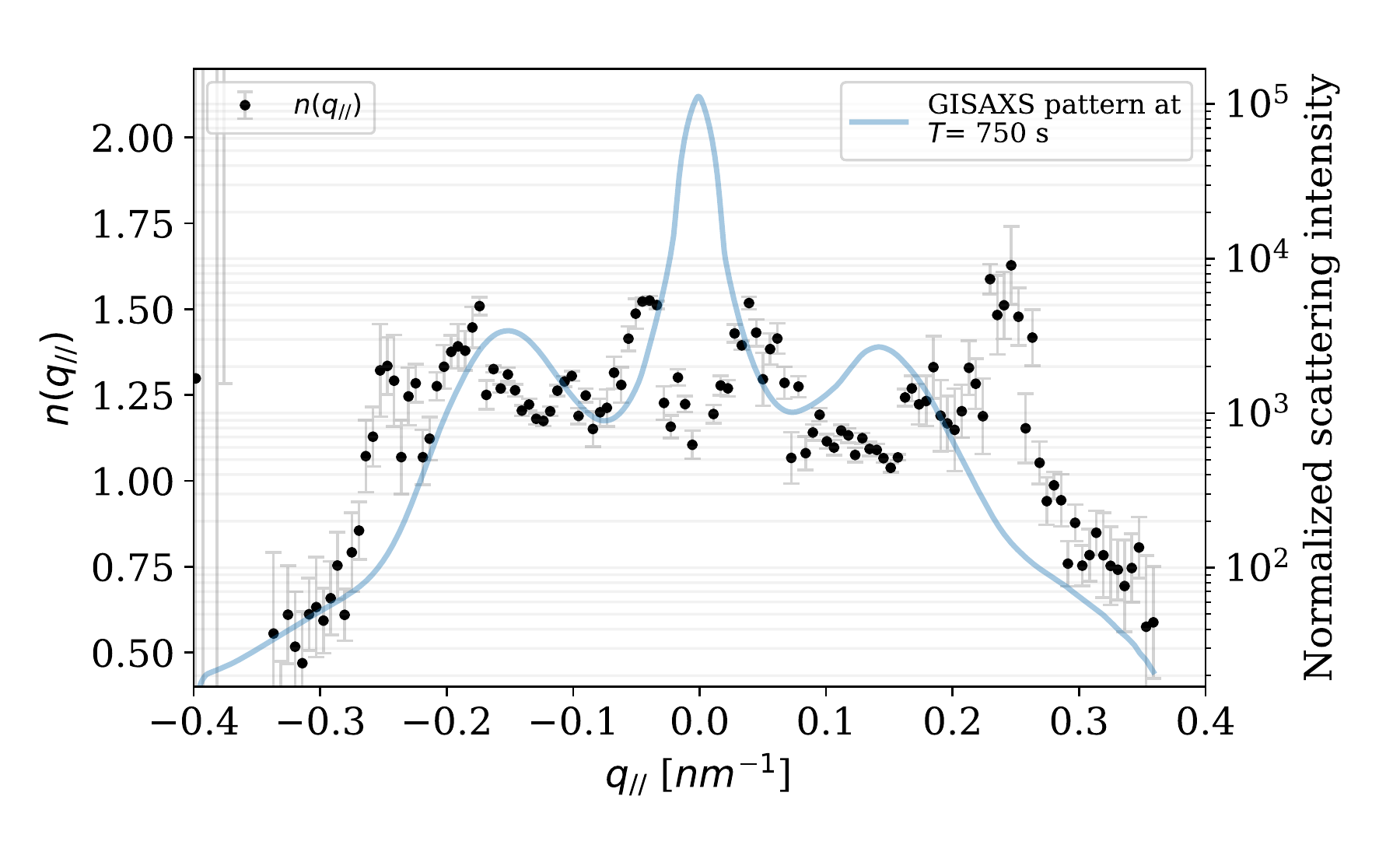}%
\includegraphics[width=3.2 in]{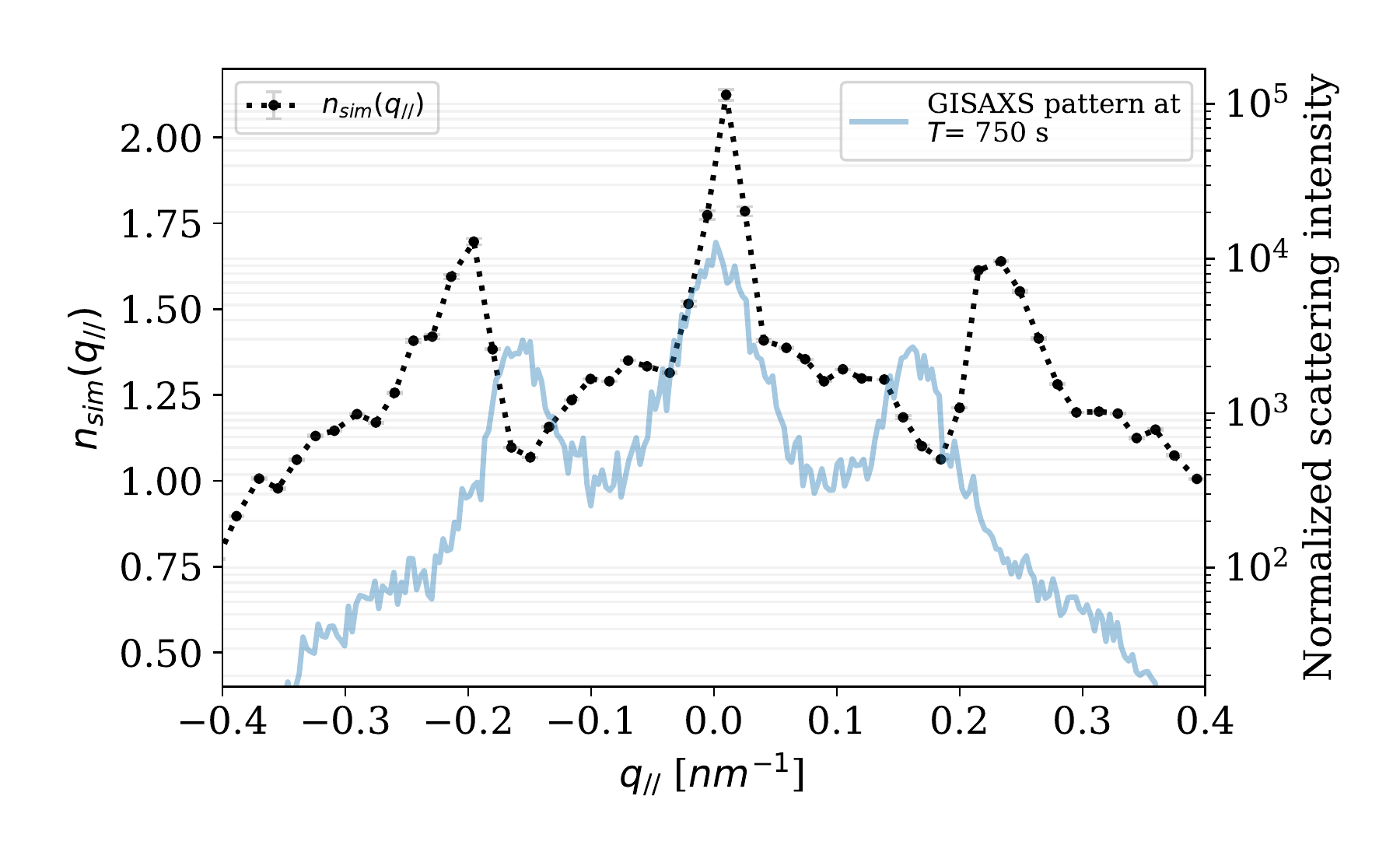}%
\caption{\label{fig:Ar_g2_and_sil} $\tau(q_{||})$ and  $n(q_{||})$ during the late stage of Ar$^{+}$ patterning of Si. Left figures: Measurements. Right figures: Simulation results.}
\end{figure*}
Plots of experimental $\tau(q_{||})$ and $n(q_{||})$ are shown in Fig. \ref{fig:Ar_g2_and_sil}. The trends seen in Fig. \ref{fig:TT_slice_allq_Ar} are confirmed and extended.  Now it can be observed that the correlation time $\tau_{||}$ is asymmetric, being higher at $+q_0$ than at $-q_0$,  This is the opposite direction of the relative peak intensities.  In addition, it's seen that there is also a peak in $\tau(q_{||})$ at the harmonic wavenumbers $\pm 2q_0$.  The peak of $\tau(q_{||})$ appears to be relatively more pronounced at the harmonic wavenumbers than does the corresponding peak in the scattering $I(q_{||})$ itself.  Near the primary correlation peaks $\pm q_0$, $n(q_{||}) > 1$, so that the relaxation is compressed exponential, as noted before.  At higher values of $q_{||}$, $n$ decreases to below one, indicative of stretched exponential behavior.

The behavior of the simulations, also shown for comparison in Fig. \ref{fig:Ar_g2_and_sil}, shows generally similar trends as experiment, as discussed below.  In addition, it appears that there may be a shoulder on $n(q_{||})$ near the position of the harmonic peaks $\pm 2q_0$.

\section{Discussion}

The final surface slope distribution clearly exhibits asymmetry with preferential tendency toward a particular slope value, especially on the positive slope side (Fig. \ref{fig:AFM}). The simulation slope distribution (Fig. \ref{fig:Simulation_lattice}) is also asymmetric, though more compact.  While the structure does not reach a highly defined sawtooth stage during the period of these experiments, it appears to be moving toward such structure. Consistent with the developing ripple asymmetry, both the experiment and simulation show that asymmetries develop in the intensities of the ripple correlation peaks at $\pm q_{||}$ (Figs. \ref{fig:sq_Ar} and \ref{fig:Simulation_sq}).  The asymmetry is more prominent for the experiments than for the simulations.  However, peaks at the ripple wavenumber $\pm q_0$ in the simulated speckle-averaged intensity are sharper in the simulation, with the harmonic peaks being significantly more pronounced. This reflects that the simulated lattice looks more ordered (Fig. \ref{fig:Simulation_lattice}) compared to the experiment (Fig. \ref{fig:AFM}).

The experiments and simulations show rich structure in the development of the correlation dynamics as seen in the parameters $\tau(q_{||})$ and $n(q_{||})$.  For $\tau(q_{||})$, on length scales of the ripple structure, local structure is becoming ever more long-lived as coarsening progresses (Figs. \ref{fig:TT}, \ref{fig:TT_slice_selectq} and \ref{fig:TT_slice_allq_Ar}).  For a total patterning time of T = 700 s, the correlation time for local ripple structure is about 240 s.  This qualitative behavior does not come as a surprise.  However, actual experimental confirmation of the evolving longevity is rare and, to our knowledge, the ability of coherent x-ray scattering to quantify the behavior is currently unique outside of the specialized environment of FIB/SEM instruments.

Figures \ref{fig:TT_slice_allq_Ar} and \ref{fig:Ar_g2_and_sil} show that, near the peak wavenumbers $\pm \; q_0$, the scattering intensity initially grows much more rapidly than does the correlation time $\tau$, but that at later times the intensity grows only slowly while $\tau$ continues to grow significantly. Eventually $\tau(q_{||})$ develops a peak on length scales corresponding to the ripple wavelength.  There is a secondary peak in relaxation times at the harmonic wavenumber $\pm 2q_0$ of the ripples.  It’s noteworthy that, just as the peaks in the simulation intensity at $\pm q_0$ are sharper than in experiment, so too the peaks in $\tau(q_{||})$ are sharper in the simulation.  There are also differences in behavior of the experiment and simulations for $\tau(q_{||})$ near the origin $q_{||} = 0$.  However, there are reasons for additional care in trusting results in this range because, in the experiment, the scattered x-rays may be mixing with the tails of the specular beam and, in the simulations, finite size effects presumably become important.  

The $\tau(q_{||})$ values are noticeably higher on the positive side of the $q_{||}$ axis, particularly in the experimental results.  Presumably this reflects asymmetry in the dynamic processes on the two sides of the ripples.  The HPB theory predicts rich dynamics on the terraces of the late-stage sawtooth structures \cite{harrison2017emergence}, so this may be related.  However, we have not been able to construct a simple model explaining the asymmetry.

Summarizing comparison of ripple structure, intensities and correlation times, while there is more order to the ripple pattern in the simulations than in the experimental results, there is more asymmetry in the experiment as observed in both the speckle-averaged intensity $I(q_{||})$ and $\tau(q_{||})$.  It should be noted that there is uncertainty in the coefficients of the terms entering the HPB simulations because of uncertainty in $Y(\theta)$ and no attempt was made to vary the coefficients in an \textit{ad hoc} manner to seek better agreement.  Moreover, no attempt was made to include any effects associated with initial surface structure \cite{munoz2012independence, kim2013role}, though we expect those to be small since experiments started with a polished Si wafer.

Turning to the relaxation exponent $n(q_{||})$, at early times the fluctuation relaxation processes are consistent with being simple exponential in nature (i.e. exponent \textit{n} = 1), as expected for linear theory behavior.  As patterning continues, however, Figs. \ref{fig:TT_slice_selectq} and \ref{fig:TT_slice_allq_Ar} show that the relaxation exponents $n(q)$ evolve in both the experiments and the simulations, with the system exhibiting compressed exponential relaxation on length scales comparable to or longer than that of the ripples and stretched exponential relaxation on much shorter length scales.  The simulations show clear peaks in $n(q_{||})$ at slightly higher $|q_{||}|$ than the intensity peak positions $\pm q_0$.  These are less clear in the experimental results, but those results are also suggestive, especially on the positive side of the $q_{||}$ axis.  Why relaxation exponents should peak there is unknown.

As we noted in Ref. \cite{myint2021gennes}, a common feature of nonlinear models of ion beam nanopatterning is the inclusion of the Kardar-Parisi-Zhang (KPZ) quadratic nonlinearities $h_x^2,h_y^2$.   For small $\Delta t$, simulations of the KPZ model are well fit with compressed exponential behavior \cite{mokhtarzadeh2017simulations} and the leading terms in the KPZ model dynamics \cite{katzav2004numerical} suggest an effective exponent $n \approx (2+2\alpha)/z \approx 1.74$, where $\alpha$ and $z$ are the roughness and dynamic exponents respectively for (2+1) dimensional growth. Thus, the KPZ effective compressed exponent $n$ for small $\Delta t$ is approximately equal to that observed at the nanoripple wavenumber peaks $\pm q_0$ in the present experiments.  It is unknown whether the HPB equation is in the KPZ universality class at long length scales, though the relaxation exponents found here appear to be similar.  As we noted previously, the inclusion of lower order terms \cite{makeev2002morphology} in the minimal Kuramoto-Sivashinsky (aKS) equation reproduces some of the kinetics features observed here, and the aKS equation is known to exhibit KPZ dynamics at large length scales.  This could potentially provide a clear connection with the KPZ behavior.

As we also discuss in Ref. \cite{myint2021gennes}, the strong peak in intensity and correlation times at the ripple wavenumbers $\pm q_0$ is reminiscent of de Gennes narrowing in liquids.  Moreover, as discussed there, compressed exponential decay observed here on length scales comparable to the ripple wavelength suggests the absence of short decay times and may be related to the concept of structural persistence. In contrast, the lower exponents observed at longer and shorter length scales suggests exponential or even stretched exponential behavior at early times, possibly indicating the lack of such persistence.  Moreover, compressed exponential behavior at short times $\Delta t$ in both soft materials \cite{cipelletti2005slow} and metallic glasses \cite{ruta2012atomic} has been attributed to collective ballistic flow of local structures due to internal stress relaxation. It's notable that some theoretical approaches to understanding ion beam nanopatterning use fluid dynamic models with stress relaxation as a driving force \cite{castro2012hydrodynamic,castro2012stress,norris2012stress,moreno2015nonuniversality,munoz2019stress}. These might provide a direct connection between the compressed exponential behavior of ion beam nanopatterning observed here and that observed in glasses.  This would be an attractive direction for future study.

\begin{acknowledgments}
We thank Andreas Mutzke for providing the SDTrimSP simulation program, R.M. Bradley for discussions and S. Norris for help with the PyCraters library. We also thank Josh Bevan (Boston University Research Computing Services) for help with optimizing the numerical simulations and our reviewers for their constructive comments. 

This material is based on work partly supported at BU by the National Science Foundation (NSF) under Grant No. DMR-1709380. X.Z. and R.H. were partly supported at UVM by the U.S. Department of Energy (DOE) Office of Science under Grant No. DE-SC0017802. Experiments were done at the Coherent Hard X-ray (CHX) beamline at National Synchrotron Light Source II (NSLS-II), a U.S. Department of Energy (DOE) Office of Science User Facility operated for the DOE Office of Science by Brookhaven National Laboratory under Contract No. DE-SC0012704. The custom UHV sample holder, designed by P.M. and K.F.L, was built at Scientific Instrumentation Facility (SIF) at Boston university. For the AFM images, Bruker Dimension 3000 Atomic Force Microscope at Precision Measurement Laboratory at the Boston University Photonics Center was utilized.

\end{acknowledgments}

\bibliography{ionpatterning_references.bib}

\end{document}